\documentstyle[amssymb,aps,twocolumn,psfig,prb,tighten]{revtex}
\newcommand{\U}{UPd$_2$Al$_3$~}
\newcommand{\la}{\langle}
\newcommand{\ka}{\kappa}
\newcommand{\ra}{\rangle}
\newcommand{\vq}{\vec{q}}
\newcommand{\vk}{\vec{k}}
\newcommand{\vQ}{\vec{Q}}
\newcommand{\tJ}{\tensor{J}}
\newcommand{\ua}{\uparrow}
\newcommand{\da}{\downarrow}
\newcommand{\al}{\alpha}
\newcommand{\be}{\beta}

\begin{document}
\draft 
\title{Dual model for magnetic excitations and superconductivity in \U} 
\author{P. Thalmeier}

\address{Max-Planck-Institute for Chemical Physics of Solids\\ 
D-01187 Dresden, Germany} 
\date{\today} 
\maketitle 

\begin{abstract}
The Heavy Fermion state in \U may be approximately described by
a dual model where two of the three U- 5f electrons are in a localized
state split by the crystalline electric field into two low lying
singlets with a splitting energy $\Delta\simeq$ 6 meV. The third 5f electron 
has itinerant character and forms the Heavy Electron
bands. Inelastic neutron scattering and tunneling experiments suggest
that magnetic excitons, the collective propagating crystal field
excitations of the localized 5f electrons, mediate superconducting (sc)
pairing in \U. A theory for this novel mechanism is developed within a
nonretarded approach. A model for the magnetic
exciton bands is analyzed and compared with experiment. The sc pair
potential which they mediate is derived and the gap
equations are solved. It is shown that this mechanism
favors an odd parity state which is nondegenerate due to the
combined symmetry breaking by the crystalline electric field and the AF 
order parameter. A hybrid model including the spin fluctuation contribution
to the pairing is also discussed.
\end{abstract}

\vspace{0.5cm}

\section{Introduction}

Among the uranium- based Heavy Fermion (HF) superconductors \cite{Sigrist91}
(sc) unconventional behaviour is the rule rather than the
exception. In the canonical intermetallic U- compounds one observes
orbitally degenerate sc states with associated double
transition\cite{Fisher84} and two upper critical field curves as in
the case of UPt$_3$ \cite{Bruls90}, furthermore coexistence and
competition of different unconventinal sc states with a SDW- like
state in the U$_{1-x}$,Th$_x$Be$_{13}$ was reported \cite{Kromer98}. In
addition URu$_2$Si$_2$ shows superconductivity embedded in a
region with a yet undetermined hidden order parameter \cite{Amitsuka99}. In all
these cases, when antiferromagnetism (AF) is observed it has no well
defined long range order and is characterized by a very small moment
size of order 10$^{-2}\mu_B$ or less. The situation is distinctly different in
the compound \U \cite{Geibel91}. Its normal state is characterized by
a specific heat $\gamma$- value of 120 mJ/mole K$^2$ much smaller than
observed in the previous compounds, it is therefore of only moderate
HF character and is superconducting below T$_c$=1.8 K. On the
other hand it has long range AF order below T$_N$=14.3 K with much larger
moments of almost atomic like size with $\mu=0.85\mu_B$. This indicates
that in addition to the itinerant electrons there must be nearly
localized 5f- electrons present. They result from the 5f$^2$
configuration of the U$^{4+}$ ionic species which is the dominating
one in this compound \cite{Grauel92}. This is supported by the
temperature dependent susceptibility which shows an extremely strong
hexagonal ac- anisotropy with an almost constant $\chi_c$ and a
much larger $\chi_a(T)$ which is strongly temperature dependent with a
typical behaviour known from crystalline electric field (CEF)- split
localized 5f states with a singlet- (nonmagnetic) ground state and a
first excited state at an energy estimated as $\Delta\simeq$6 meV. Knight
shift analysis in the normal state obtained from $\mu$sr- experiments
confirm the presence of localized 5f moments \cite{Feyerherm94}. Since
the CEF ground state is a singlet the AF order must be of the induced
moment type, i.e. due to mixing with the first excited CEF state caused by the
inter- site exchange. 

The most direct confirmation for this dual
nature of 5f- electrons in \U is obtained from inelastic neutron
scattering experiments which have been able to identify the collective
propagating excitations that originate in CEF excitations of energy
$\Delta$ and are broadened into a dispersive band usually termed 'magnetic
excitons' due to the action of the intersite exchange. These modes
extend up to 10 meV and for a wave vector along the hexagonal axis are
quite sharp and well defined \cite{Mason97}. More recently it was
found \cite{Bernhoeft98,Sato97,Bernhoeft00} that at the AF wave vector
$\vQ=(0,0,\pi)$ in reduced units (App. A) a strong
interaction of these collective modes of localized moments with the
heavy conduction electrons exists which leads to a resonance- like
structure in the dynamical structure function due to a near degeneracy
of the exciton mode energy at $\vQ$ and the superconducting
gap. Before that tunneling measurements \cite{Jourdan99} have already shown
that strong coupling anomalies in the tunneling current exist in the
same energy range where the magnetic excitons at $\vQ$ were
found. This has lead to the conclusion that they are indeed
the bosonic 'glue' which is responsible for the formation of Cooper
pairs in this compound \cite{Sato01}. The argument is quite
analogous to the strong coupling conventional electron-phonon
superconductors where the phonon spectrum known from inelastic neutron
scattering leaves its imprint on the sc tunneling spectrum. \U is
therefore the first case of an unconventional HF superconductor where
a similar comparison has lead to the identification of the
pairing mechanism, in this case the pairing is mediated by the
exchange of magnetic excitons.

In the present work this novel type of pairing mechanism is
investigated in detail. It is different from both the electron- phonon
mechanism in conventional superconductors and from the
spin fluctuation mechanism which is commonly assumed to be present in
the unconventional superconductors with strongly correlated
electrons. The former is mediated by slightly damped real frequency bosons
(phonons) which do not couple to the spin degrees of freedom, thus
allowing only for spin singlet pairs. The latter is mediated by
strongly overdamped spin- fluctuations, i.e. bosons with purely
imaginary frequency which couple to conduction electron spins in a
rotationally invariant manner and in
principle allow for both spin singlet and triplet pairing.
The new pairing mechanism discussed here has distinctly different
features: It is mediated by magnetic excitons which are real
frequency, propagating bosonic modes that couple to the conduction
electron spin, however in a way which explicitly breaks spin
rotational invariance due to the presence of the CEF
splitting. Together with the effect of the AF order parameter this
will lead to a complete splitting of triplet states resulting in
nondegenerate odd parity pair states.

It is the aim of this work to study within a generic dual model of 5f
electrons (Sect. II) the magnetic exciton bands (Sect. III) and the
associated pair potential (Sect .IV) of this mechanism, incorporating  both the
localized CEF states which originate
from the 5f$^2$ configuration of the U$^{4+}$ state of \U and the
conduction electrons. Furthermore
the gap equations and their explicit solutions together with their
node line structure for a quasi one dimensional model are discussed
explicitly; it is also shown how the presence of the AF background
influences the gap structure and lifts the degeneracy (Sect .V). Finally
we discuss a hybrid model including also the usual
spin fluctuation contribution (Sect. VI) and a summary and outlook is given
in Section VII.

\section{Basic Hamiltonian for the dual 5f- model}

The pronounced susceptibility anisotropy of UPd$_2$Al$_3$ with an
almost constant $\chi_c$ and a strongly temperature
dependent $\chi_a$(T) with a maximum at T=50 K is
perhaps the most direct evidence for the presence of localised CEF-
states \cite{Grauel92}. In fact the $\chi_{a,c}$(T)-
dependence was used to extract the hexagonal CEF- scheme in
Ref.(\onlinecite{Grauel92}) where it was conluded that
U$^{4+}$(5f$^2$) ground state and first excited states are both
singlets. However later it was found \cite{Shiina01} that this cannot
explain the temperature dependence of the staggered magnetisation and
a singlet- doublet system was proposed instead. The magnetic
excitations are very similar for both CEF systems except for a small
difference in their temperature dependence. For simplicity we
therefore use the singlet-singlet system with the ground state
$|g\rangle$ (0 meV) and excited state $|e\rangle$ ($\Delta$= 6
meV). The value of $\Delta$ is obtained from a fit to the experimental
excitations (Sect.IV) and is close to the one obtained from the
temperature variation of susceptibility and AF order parameter
\cite{Shiina01}. All higher CEF levels starting at $\sim$ 10 meV will
be neglected. The total Hamiltonian, including the localized part
H$_{CEF}+H_{ff}$, the conduction band states (H$_c$) of the remaining
itinerant 5f- electron and its interaction with localized states (H$_{cf}$)
was introduced in Ref.(\onlinecite{Sato01}) as 

\begin{eqnarray}
\label{HAM}
H&=&H_c+H_{CEF}+H_{ff}+H_{cf}\nonumber\\
H&=&\sum _{\vk\sigma}\epsilon_{\vk\sigma}c^\dagger_{\vk\sigma}c_{\vk\sigma}
+\Delta\sum _i\mid e\rangle\langle e\mid _i \\
&&-\sum _{\ll ij\gg}J_{ff}(ij)\vec{J}_i\vec{J}_j\nonumber
-2I_0(g-1)\sum _i\vec{s}_i\vec{J}_i
\end{eqnarray}

Here $\epsilon_{\vk\sigma}$ is the dispersion of itinerant electrons
described by (c$_{\vk\sigma}$,$\vec{s}_i$) creation and spin
operators. The most important Fermi surface (FS) sheet has the shape of a
slightly corrugated cylinder along the hexagonal c-
axis\cite{Knoepfle96,Inada99}. It can be modeled by the expression 

\begin{equation}
\label{DIS}
\epsilon_{\vk\sigma}=\epsilon_{\perp}(\vk_{\perp}\sigma)
-2t_\parallel\cos k_z 
\end{equation}

Where $\epsilon_{\perp}(\vk_{\perp}\sigma)$ is the dispersion
perpendicular to the c-axis whose precise form is unimportant in the
following, the much smaller dispersion $\parallel$ to c which is
responsible for the corrugation of the FS cylinder is determined by an
effective hopping energy t$_{\parallel}$. The localised 5f- electrons
have a total angular momentum $\vec{J}_i$ (J=4). Within the
singlet-singlet subspace they can be represented by a pseudospin
$\vec{S}_i$ (S=$\frac{1}{2}$) with the correspondence

\begin{equation}
\label{JOP}
J_x=\alpha S_x,~ J_y=\alpha S_y, J_z=\epsilon(\frac{1}{2}-S_z)
\end{equation}

This means that $\la e|J_x|g\ra$= -i$\la e|J_y|g\ra$=
$\frac{1}{2}\alpha$ and $\la e|J_z|g\ra$ =0, i.e. inelastic
singlet-singlet transitions can only be excited by the transverse
operators J$_{x,y}$. The third term describes a possible superexchange
J$_{ff}$= 4(g-1)$^2$t$_{ff}^2$/U between the localised 5f- electrons
due to their remaining finite inter- site hopping t$_{ff}$. The last
term finally
represents the exchange interaction which couples the itinerant and
localised 5f- electrons of the dual model. The total effective inter-
site exchange between localized 5f- moments is then given by its
Fourier transform as

\begin{eqnarray}
\label{EFF}
\tJ(\vq)&=&\tJ_{ff}(\vq)+I_0^2(g-1)^2\tensor{\chi'}_0(\vq)\nonumber\\
\tensor{\chi'}_0(\vq)&=&
\tensor{\chi}_0(\vq)-\frac{1}{N}\sum_{\vq}\tensor{\chi}_0(\vq)
\end{eqnarray}

Where the second (RKKY) part is in principle determined by the static 
conduction electron susceptibility $\tensor{\chi}_0(\vq)$. However in the following $\tJ(\vq)$ will be treated as an empirical quantity to be parametrized by fitting the actually observed magnetic excitations to the theoretical predictions based on the dual Hamiltonian of Eq.(\ref{HAM}).

\section{The antiferromagnetic singlet-singlet system}

The 'induced' magnetism of singlet-singlet CEF- systems is well studied
for Pr- compounds (for a review, see
Ref.(\onlinecite{Jensen91})). Since U$^{4+}$ has also a f$^2$
electron configuration similar to Pr$^{3+}$ singlet-singlet
systems may also exist in U-compounds with rather localised 5f- states
such as \U. In a singlet-singlet system the ground state is
nonmagnetic, i.e. $\la g|\vec{J}|g\ra \equiv$ 0. Nevertheless magnetic moments may appear spontaneously at T$_N$ if the effective exchange is strong enough to mix the excited state $|e\rangle$ into the ground state  $|g\rangle$. The mixed state $|\tilde{g}\ra$ will then have nonzero moment due to the nondiagonal matrix element $\alpha$=2$\la e|J_x|g\ra$. In this case the AF transition is preceded by the softening of a 'magnetic exciton' mode at the AF wave vector ($\vQ=(0,0,\pi)$ in \U).

\subsection{Exchange model and origin of induced AF order}

The magnetic exciton mode can be interpreted as originating from local
CEF- excitations $|g\ra\leftrightarrow |e\ra$ with energy $\Delta$
which propagate from site to site due to the effective exchange and
this process leads to the magnetic exciton dispersion. It is therefore
a collective excitation of localized 5f- CEF states and may be obtained
from the second and third term in Eq.(\ref{HAM}) but taking the total
effective exchange of Eq.(\ref{EFF}) instead. Within the two singlet subspace
this leads to a Hamiltonian

\begin{figure}
\centerline{\psfig{figure=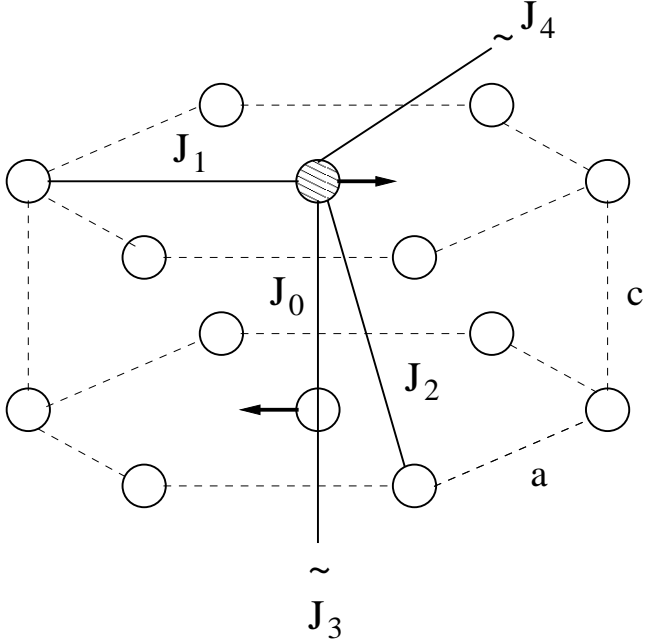,height=8.5cm,width=8.5cm}}
\vspace{1cm}
\caption{Schematic view of conventional unit cell of \U with only U-
atoms shown (a=5.35 $\AA$, c=4.185 $\AA$). Arrows indicate magnetic
moments with FM- in plane order and AF stacking along c corresponding
to $\vQ$=(0,0,$\frac{\pi}{c}$). The exchange pairs used in the model
of Sect. 3 are
also indicated. J$_3$ and J$_4$ describe the coupling to n.n.n. pairs
along c and in the ab- plane respectively. The values of J$_i$(i=0-4)
are given in the caption of Fig.4.}
\label{FIGSTR}
\end{figure}

\begin{equation}
\label{HHE}
H_0=(\Delta/\epsilon)\sum_iJ^i_z-\frac{1}{2}\sum_{\la ij\ra}
\vec{J}_i\tensor{J}_{ij}\vec{J}_j
\end{equation}

where $\tensor{J}$ is the uniaxial exchange tensor of Eq.(\ref{EFF})) with
J$^{xx}_{ij}$= J$^{yy}_{ij}$=J$^a_{ij}$ and
J$^{xx}_{ij}$=J$^c_{ij}$. Its Fourier transform  for the two AF
sublattices A,B (=$\lambda,\tau$) is given by (i$\in\lambda$)

\begin{equation}
\label{EXF}
\tJ_{\lambda\tau}(\vq)=
\sum_{j\in\tau}\tJ_{ij}\exp(-i\vq(\vec{R}_i-\vec{R}_j))
=\tJ_{\tau\lambda}(\vq)^*
\end{equation}

As mentioned in Sect.(II) the exchange functions will be empirically
parametrized to fit the magnetic exciton dispersion. The exchange
couplings included are illustrated in Fig.\ \ref{FIGSTR}. This leads to 

\begin{eqnarray}
\label{EXL} 
\tJ_D(\vq)&=&\tJ_{AA}(\vq)= \tJ_{BB}(\vq)=\nonumber\\ 
&&\tJ_1\gamma_1(\vq)+\tJ_3\gamma_2(\vq)+\tJ_4\gamma_3(\vq) \\
\tJ_N(\vq)&=& \tJ_{AB}(\vq)= \tJ_{BA}(\vq)^*=\nonumber\\ 
&&[\tJ_0+\tJ_2\gamma_1(\vq)]\gamma_0(\vq)\nonumber
\end{eqnarray}

The structure functions $\gamma_i(\vq)$ are given by

\begin{eqnarray}
\label{STR}
\gamma_0(\vq)&=&2\cos q_z \nonumber\\
\gamma_1(\vq)&=&2[\cos q_x+\cos(\frac{1}{2}q_x+\frac{\sqrt{3}}{2}q_y)+
\nonumber\\
&&\cos(\frac{1}{2}q_x-\frac{\sqrt{3}}{2}q_y)] \nonumber\\
\gamma_2(\vq)&=&2\cos 2q_z \\
\gamma_3(\vq)&=&2[\cos\sqrt{3}q_y+\cos\frac{\sqrt{3}}{2}(q_y+\sqrt{3}q_x)+
\nonumber\\
&&\cos\frac{\sqrt{3}}{2}(q_y-\sqrt{3}q_x)]\nonumber
\end{eqnarray}

where $\gamma_0(q_z\pm Q_z)= -\gamma_0(q_z)$ and $\gamma_2(q_z\pm Q_z)= 
\gamma_2(q_z)$. This leads to the important symmetries
$\tJ_D(q_z\pm Q_z)= \tJ_D(q_z)$ and $\tJ_N(q_z\pm\ Q_z)= -\tJ_N(q_z)$ 
where $\vq$ and $\vQ$=(0,0,$\pi$) are given in units of a$^{-1}$ and 
c$^{-1}$ for x,y and z components, respectively. Each of the
exchange parameters $\tensor{J}_\nu$ ($\nu$=0-4) in Eq.(\ref{EXL})
corresponds to a
given in- or out of plane neighbor shell in Fig.\ \ref{FIGSTR} is in general a
uniaxial tensor. The tensor notation will now be suppressed for
simplicity. In mean field (mf) approximation the Hamiltonian in
Eq.(\ref{HHE}) reads:

\begin{eqnarray}
\label{HMF}
H^{mf}_0&=&\sum_i[(\Delta/\epsilon)J_z(i)- h_e^\lambda J_x(i)] \nonumber\\
\vec{h}_e^\lambda &=&[J_{AA}(0)-J_{AB}(0)]\la J\ra_{\lambda}= 
J(\vQ)\la J\ra_{\lambda}\\
J_e&=&J(\vQ)=2[3J_1+3J_4+J_3-J_0-6J_2]\nonumber
\end{eqnarray}

Here $\vec{h}_e$= $\vec{h}_e(A)$= -$\vec{h}_e(B)$ =$h_e\hat{x}$ is the
staggered molecular field and $\la J\ra=\la J_x\ra = \alpha\la S_x\ra$
is the induced moment (in units of g$\mu_B$). The single- ion mf
Hamiltonian for each sublattice is then 

\begin{eqnarray}
\label{MFL}
h_0^A(i)&=&\frac{\Delta}{2}
\left(\matrix{
1&-\gamma'\cr
-\gamma' & -1\cr}\right) \nonumber\\
h_0^B(i)&=&\frac{\Delta}{2}
\left(\matrix{
1&\gamma'\cr
\gamma' & -1\cr}\right)
\end{eqnarray}

where $\gamma$'= $\gamma\la J\ra$, $\gamma$=$\frac{\alpha J_e}{\Delta}$.
The mf- energies and states are then given by

\begin{eqnarray}
\label{MFS}
\epsilon_{\pm}&=&\pm\frac{\Delta}{2}[1+\gamma '^2]^{\frac{1}{2}}\nonumber\\
|+\ra&=&u|e\ra -v|g\ra \\
|-\ra&=&u|e\ra +v|g\ra \nonumber
\end{eqnarray}

Here u=$\cos\theta$, v=$\sin\theta$ determine the rotation to the $|\pm\ra$ mf
eigenstates with 2$\theta =\tan^{-1}(\gamma')$, furthermore 
$\sin 2\theta=(1+\gamma'^2)^\frac{1}{2}$ and 
$\cos 2\theta=\gamma'(1+\gamma'^2)^\frac{1}{2}$ 
where the convention $\theta=\theta_A=-\theta_B$ has been used. From
the mf- single ion partition function Z=$\cosh\beta\epsilon$ one
obtains the mf- equations for the singlet-singlet splitting
$\Delta$'(T) in the ordered state

\begin{eqnarray}
\label{DEL} 
\hat{\Delta}(T)&=&\frac{\Delta'(T)}{\Delta}=
\frac{1}{\Delta}(\epsilon_+-\epsilon_-)=(1+\gamma'^2)^\frac{1}{2}\nonumber\\
\hat{\Delta}(T)&=&\xi\tanh [(\frac{\beta\Delta}{2})\hat{\Delta}(T)]=
2\xi\la S\ra
\end{eqnarray}

Here $\la S_z\ra\equiv\la S\ra$ is the pseudospin expectation value
which can be replaced by $\frac{1}{2}$ since T$\ll\Delta$, furthermore
$\hat{\Delta}(0)=\frac{1}{2}\alpha\gamma=\xi$. The control parameter
of the singlet- singlet system is

\begin{equation}
\label{XSI}
\xi=\frac{\alpha^2J_e}{2\Delta}
\end{equation}

which characterizes the strength of exchange vs. size of CEF splitting. It determines the AF transition temperature where $\hat{\Delta}(T_N)$=0 as

\begin{eqnarray}
\label{CRI}
T_N&=&\frac{\Delta}{2\tanh^{-1}(\frac{1}{\xi})}=
\frac{\Delta}{2\tanh^{-1}(\frac{2\Delta}{\alpha^2J_e})} \nonumber\\
\frac{1}{\xi}&=&\tanh\frac{\Delta}{2T_N}
\end{eqnarray}

The staggered induced moment as a function of temperature is given by

\begin{eqnarray}
\label{MOM}
\la\hat{J}\ra_T&=&\frac{\la J\ra_T}{\la J\ra_0}=
(\xi^2-1)^{-\frac{1}{2}}[\hat{\Delta}^2(T)-1]^{\frac{1}{2}}\nonumber\\
\la J\ra_0&=&\alpha\la S_x\ra_0=
\frac{1}{2}\alpha\frac{1}{\xi}(\xi^2-1)^{\frac{1}{2}}
\end{eqnarray}

where $\la J\ra_0$ is the saturation moment. To obtain a spontaneous induced moment $\la J\ra_0 >0$ the control parameter $\xi$ must fulfil  $\xi>\xi_c\equiv 1$. For $\xi =1+\delta$ ($\delta\ll 1$) slightly above its critical value $\xi_c =1$ the saturation moment is given by

\begin{equation}
\label{IND}
\la J\ra_0 =\frac{\alpha}{\sqrt{2}}\delta^\frac{1}{2};\; 
\delta=2\exp(-\frac{\Delta}{T_N})
\end{equation} 

\begin{figure}
\centerline{\psfig{figure=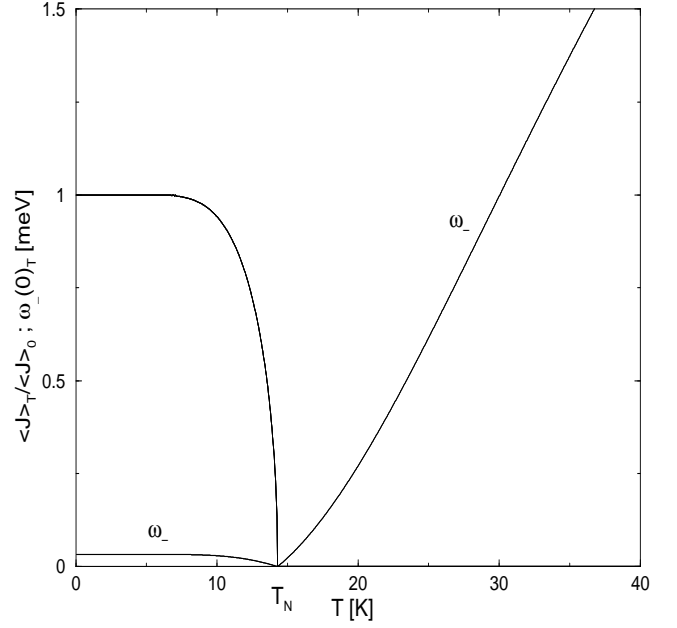,height=8.5cm,width=8.5cm,angle=-90}}
\vspace{1cm}
\caption{Temperature dependence of magnetic exciton soft mode
frequency $\omega_-$(0)$_T$ (thin line) and normalized induced AF 
moment $\la J\ra_T/\la J\ra_0$ (thick line) as given by Eq.(\ref{MOM})
The small finite $\omega_-$(0)$_T$ below T$_N$ is due to exchange
anisotropy. Physical parameters are the same as in Fig.4.}
\label{FIGIND} 
\end{figure}

Therefore when $\xi\geq 1$ or T$_N\ll\Delta$ the saturation moment
becomes exponentially small. This distinguishes an induced moment
system from a conventional magnet with degenerate ground state where
the saturation moment is a constant independent of T$_N$. Note that the above relations
also hold in the ferromagnetic case if we replace J$_e\rightarrow
$J$_{ff}$(0)= J$_{AA}$(0)+J$_{AB}$(0). The temperature dependence of
the induced moment is shown in Fig.\ \ref{FIGIND}

\subsection{Magnetic exciton dispersion and AF soft mode}

The local CEF excitation at site i with energy $\Delta$ can propagate
through the effect of the inter- site exchange whose transverse parts
lead to transitions $|e\ra_i\rightarrow|g\ra_i$ and
$|g\ra_j\rightarrow|e\ra_j$ simultaneously. This causes a widening
of $\Delta$ into a dispersive band of 'magnetic excitons'. It should
be stressed that these elementary excitations do {\em not} require the
presence of a
magnetic order parameter as in the case of spin waves. Magnetic
excitons exist already in the paramagnetic (PM) phase. As function of
temperature they may exhibit softening at a certain wave vector which
signifies a magnetic transition that leads to the appearance of
induced moments. This mechanism is well studied for Pr
metal\cite{Jensen91} and some 
compounds where the ordering wave vector may be incommensurate in
contrast to the commensurate AF wave vector $\vQ$ in \U. In the
following a detailed description of the magnetic exciton modes for \U
is given since they are the bosonic excitations thought to mediate
superconductivity in this compound  (Sect. IV.B). The calculation is done
most conveniently in the linear response formalism in RPA which starts from the
dynamical susceptibility of the singlet-singlet system \cite{Jensen91}.

\begin{equation}
\label{CHI}
\chi_{ij;\alpha\beta}^{\lambda\mu}=
-\la T\{J_{i\alpha}^\lambda(\tau)J_{j\beta}^\mu(0)\}\ra
\end{equation} 

where i,j= lattice site, $\lambda,\mu$=  AF sublattices (A,B) and
$\alpha,\beta$= x,y,z are cartesian coordinates. Its Fourier transform
is given in RPA by 

\begin{equation}
\label{RPA}
\tensor{\chi}(\vq,\omega)=[1-\tensor{u}(\omega)\tJ(\vq)]^{-1}
\tensor{u}(\omega)
\end{equation}

Here $\tensor{\chi}$, $\tensor{u}$ and $\tJ$ are 4$\times$4 tensors,
e.g. $\tensor{\chi}$=$\{\chi\}_{\lambda\alpha,\mu\beta}$ with
$\lambda,\mu$= A,B and $\alpha,\beta$=x,y transverse cartesian
coordinates. The tensor $\tensor{u}$ is diagonal in $\lambda,\mu$ and
$\tensor{J}$ is diagonal in $\alpha,\beta$. The expression for
$\tensor{u}(\omega)$ is given in App.(B). To keep it in a simple form
in the ordered state ($\la J\ra >0$) a real space back rotation by
an angle $\theta_r$= $\theta^A_r$= -$\theta^B_r$ has to be applied to
compensate for the effect of the molecular field. Under this rotation
the exchange tensors transform as

\begin{equation}
\label{JPR}
\tJ'(\lambda\mu)=\tensor{D}(\theta_r^\lambda)\tJ(\lambda\mu)
\tensor{D}^T(\theta_r^\mu)
\end{equation}

The magnetic exciton modes are then obtained as poles of the
dynamical susceptibility in Eq.(\ref{RPA}) leading to the secular equation 
$\det\tensor{\chi}(\vq,\omega)$=0 or

\begin{equation}
\label{SEC}
|\tensor{u'}^{-1}(\omega)-\tJ'(\vq)|=0
\end{equation}

With $\tensor{u'}^{-1}(\omega)$ given in Eq.(\ref{URO}). The magnetic exciton
branches obtained from the secular equation above will now be
discussed for several cases. For brevity the primes in the transformed
tensors will be suppressed unless explicitly needed.

\subsubsection{paramagnetic phase, isotropic exchange}

In this case there are no sublattices A,B and only one exchange
function $\tJ(\vq)=\tJ_D(\vq)+\tJ_N(\vq)$ appears, furthermore
$\Delta'=\Delta$ holds and Eq.(\ref{SEC}) then yields the PM
exciton dispersion

\begin{eqnarray}
\label{DI1}
\omega(\vq)&=&\Delta-\alpha^2\la S\ra J(\vq) \nonumber\\
\omega(\vq)&=&\Delta[1-\frac{\alpha^2J(\vq)}{2\Delta}
\tanh\frac{\beta}{2}\Delta]
\end{eqnarray}

At the AF point $\vQ$ (PM zone boundary) we have J($\vQ$)=
J$_D$($\vQ$)+J$_N$($\vQ$)$\equiv$ J$_e$ and therefore

\begin{equation}
\label{DI2}
\omega(\vQ,T)=\Delta[1-\xi\tanh\frac{\beta}{2}\Delta]
\end{equation}

This zone boundary mode in the paramagnetic BZ (PMBZ) becomes soft
when T$\rightarrow$ T$_N$ from above according to Eq.(\ref{DI2}).

\begin{equation}
\label{DI3}
\omega(\vQ,T)=\frac{1}{2}(\frac{\Delta}{T_N})^2(\xi-\frac{1}{\xi})
(T-T_N)
\end{equation}

It approaches zero linearly above T$_N$. Below T$_N$ the induced moment
of Eq.(\ref{MOM}) appears staggered along the c- axis leading to two AF
sublattices A,B. Then $\vQ=(0,0,\pi)$ becomes the new center of the
antiferromagnetic BZ (AFBZ). One has the connection $\vq~'=\vq-\vQ$,
where $\vq\in$ PMBZ ($|q_z|\leq\pi$) and $\vq~'\in$ AFBZ 
($|q'_z|\leq\frac{\pi}{2}$) as illustrated in the inset of Fig.\
\ref{FIGOPA}. The prime for the AFBZ will be suppressed i.f. unless
explicitly necessary.

\subsubsection{AF induced moment phase, isotropic exchange}
 
For the solution of Eq.(\ref{SEC}) one now needs the rotated exchange tensor
in Eq.(\ref{JPR}). For isotropic $\tJ(\lambda\mu)$ one obtains

\begin{eqnarray}
\label{JPS}
\tJ'(\lambda\mu) &=&\tJ(\lambda\mu)\tensor{D}(\theta_r^\lambda)
\tensor{D}^T(\theta_r^\mu) \nonumber\\
\tJ'_D(\vq)&=&\tJ_D(\vq)\\
\tJ'_N(\vq)&=&\frac{1-\gamma'^2}{1+\gamma'^2}\tJ_N(\vq)
\nonumber
\end{eqnarray}

Where $\tJ'_D(\vq), \tJ'_N(\vq)$ are the intra- and inter- sublattice
exchange tensors in the rotated (x',y,z') coordinate system
respectively. From
Eq.(\ref{SEC}) one then obtains the magnetic exciton dispersion

\begin{eqnarray}
\label{DI4}
\omega_\mp^2(\vq)&=&[\Delta'-\alpha^2\la S\ra (J_D(\vq)\mp rJ_N(\vq))]\\
&&[\Delta'-\alpha^2\la S\ra (J_D(\vq)\mp J_N(\vq))] \nonumber\\
r&=&\frac{1-\gamma'^2}{1+\gamma'^2}=\frac{2-\xi^2}{\xi^2}
\nonumber
\end{eqnarray}

Here the last identity holds because T$\ll\Delta$ where $\gamma'^2$=
$\hat{\Delta}^2-1\simeq\xi^2-1$. Equivalently this may be written

\begin{eqnarray}
\label{DI5}
\omega_\mp^2(\vq)&=&[\Delta'-\alpha^2\la S\ra J_\pm(\vq)]^2-
(\alpha^2\la S\ra)^2\gamma'^4\tilde{J}_N^2(\vq)\nonumber\\
\tilde{J}_N(\vq)&=&(1+\gamma'^2)^{-1}J_N(\vq);\;
J_\pm(\vq)=J_D(\vq)\pm\tilde{J}_N(\vq)
\end{eqnarray}

\begin{figure}
\centerline{\psfig{figure=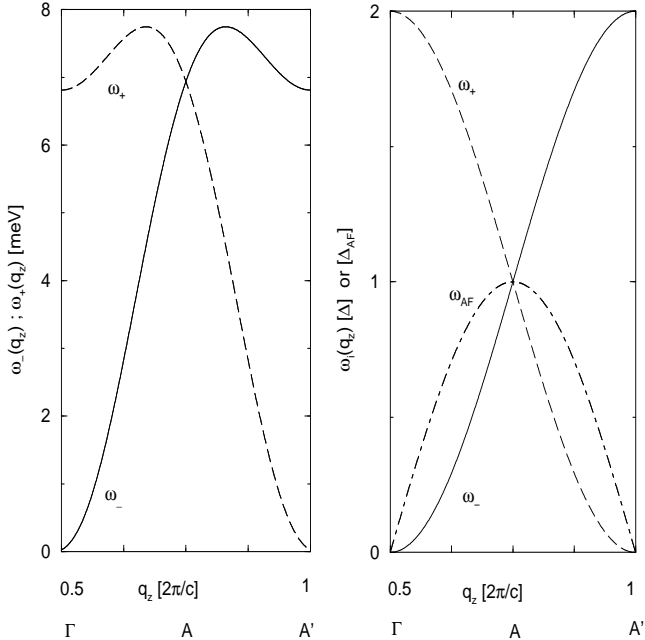,height=8.5cm,width=8.5cm,angle=-90}}
\vspace{1cm}
\caption{{\em left panel}: Magnetic exciton branches $\omega_\pm$ in the
AFBZ ($\Gamma$AA' is equivalent to A$\Gamma$A, see also inset of
Fig. 8. Physical parameters as in Fig.4. {\em right panel}:
Simplified model with only n.n exchange along c. Comparison of
magnetic exciton branches $\omega_\pm$ for the CEF- dominated case
($\xi\geq$1) and (twofold degenerate) AF spin wave dispersions
$\omega_{AF}$ for the
exchange dominated case ($\xi\gg$1), normalization to $\Delta$ and
$\Delta_{AF}$ of Eq.(\ref{DI9}) respectively}
\label{FIGDI1} 
\end{figure}

In the AFBZ $\omega_-(\vq)$ and  $\omega_+(\vq)$ correspond to
acoustic (A) and optical modes (O) respectively. Due to the property
J$_N(\vq\pm\vQ)$=-J$_N(\vq)$ the optic mode  $\omega_+(\vq)$ with
$\vq\in$ AFBZ is the downfolded (shifted by an AF reciprocal lattice
vector $\vQ$) acoustic mode  $\omega_-(\vq)$ with $\vq\in$ PMBZ. This
connection is illustrated in Fig.\ \ref{FIGDI1} (left panel). It is
instructive to consider two limiting cases for this dispersion:\\

\noindent
(i) $\xi\geq$ 1 (r$\rightarrow$ 1)\\
In this marginally critical case r$\geq$1 and Eq.(\ref{DI4}) reduces to 

\begin{equation}
\label{DI6}
\omega_\mp^2(\vq)=\Delta'-\alpha^2\la S\ra J_D(\vq)
\mp\alpha^2\la S\ra J_N(\vq)
\end{equation}

The acoustic mode $\omega_-(\vq)$ evolves from the paramagnetic soft
mode $\omega(\vq)$ of Eq.(\ref{DI1}) and the optic mode  $\omega_+(\vq)$ is
obtained by downfolding  $\omega_-(\vq)$. Therefore we have two
strongly split A,O modes in the AFBZ in the present case. Including
only n.n. exchange along the c-axis in Eq.(\ref{EXL}) leads to J$_D$=0 and
J$_N(\vq)$=2J$_0\cos q_z$ and we obtain with $\alpha^2\la S\ra
=\xi\Delta\simeq\Delta$:

\begin{equation}
\label{DI7}
\omega_\mp(\vq)=\Delta(1\pm\cos q_z)
\end{equation}

The two modes are periodic in the reduced AFBZ.\\ 

\noindent
(ii) $\xi\gg$ 1 (r$\rightarrow$ -1)
In this limit the magnetic excitations are exchange dominated and the
influence of the CEF is negligible, therefore they can essentially be
viewed as AF spin waves. One obtains from Eq.(\ref{DI4}):

\begin{eqnarray}
\label{DI8}
\omega_\mp^2(\vq)&=&[\Delta'^2-(\alpha^2\la S\ra)^2J_N^2(\vq)]^\frac{1}{2}
\nonumber\\
&=&\alpha^2\la S\ra)[J_N^2(0)-J_N^2(\vq)]^\frac{1}{2}
\end{eqnarray}

where we used $\Delta'=\Delta\xi=\alpha^2\la S\ra_0J_e= \alpha^2\la
S\ra J_N(0)$. Again, for only n.n. exchange along the c- axis (J$_D$=0,
J$_N(\vq)$=zJ$_0\cos q_z$, z=2) one obtains

\begin{equation}
\label{DI9}
\omega_{AF}(\vq)=\omega_\mp(\vq)=\alpha^2\la S\ra zJ_0|\sin q_z|=
\Delta_{AF}|\sin q_z|
\end{equation}

which is the spinwave dispersion of a simple
AB-antiferrromagnet. Due to isotropic exchange it is twofold
degenerate by which it is clearly distinguished from the strongly
split A-O exciton modes in the marginally critical ($\xi\geq$ 1) case. An
illustrative comparison of the two cases ($\xi\geq$ 1, $\xi\gg$ 1) is
shown in Fig.\ \ref{FIGDI1} where $\Delta_{AF}$= $\alpha^2\la S\ra zJ_0$ is
the spin wave band width with $\Delta_{AF}/\Delta=\xi\gg$1

\subsubsection{AF phase, anisotropic exchange}

In the isotropic case a gapless Goldstone mode for $\vq\rightarrow$ 0
(AFBZ) will be present for any $\xi\geq$1 above the critical value. However due to the uniaxial symmetry of the effective exchange
tensor Eq.(\ref{EFF}) in principle one has to expect gapped magnetic
excitations. For simplicity we assume that the anisotropy is
independent of the spin pairs, then we may define uniaxial
$J_{D,N}^a(\vq)$, $J_{D,N}^c(\vq)$ exchange functions in obvious
notation where a,c refers to the hexagonal directions. The magnetic
exciton modes may again be obtained as solutions of Eq.(\ref{SEC}), however
the transformed $\tJ'_D(\vq)$ and $\tJ'_N(\vq)$ are no longer
proportional to the unit matrix as in Eq.(\ref{JPS}). Instead one has for the
restricted (x',y) coordinate system:

\begin{equation}
\label{TED}
\tJ'_D=\tensor{D}(\theta_r^A)\tJ_D\tensor{D}^T(\theta_r^A)=
\left(\matrix{
u_r^2J_D^a+v_r^2J_D^c & 0\cr
0                     & J_D^a\cr}\right)
\end{equation}

\begin{equation}
\label{TEN}
\tJ'_N=\tensor{D}(\theta_r^A)\tJ_N\tensor{D}^T(\theta_r^B)=
\left(\matrix{
u_r^2J_N^a-v_r^2J_N^c & 0\cr
0                     & J_N^a\cr}\right)
\end{equation}

where u$_r=\cos\theta_r$, v$_r=\sin\theta_r$ and the rotation angles
for the two sublattices are $\theta_r^A=\theta_r=-2\theta$ and
$\theta_r^B=-\theta_r=2\theta$. Therefore (App. B) u$_r=\cos
2\theta = a= (1+\gamma'^2)^{-\frac{1}{2}}$ and v$_r=\sin 2\theta =b= 
\gamma'$a. From Eq.(\ref{SEC}) and using Eq.(\ref{TED},\ref{TEN}) we then
obtain the magnetic exciton modes in the most general case considered
here as

\begin{eqnarray}
\label{DIA}
\omega_\pm^2(\vq)&=&[\Delta'-\alpha^2\la S\ra J_\pm(\vq)]^2
-(\alpha^2\la S\ra)^2\gamma'^4\tilde{J}(\vq)_\pm^2 \nonumber\\
J_\pm(\vq)&=&(J_D\pm J^{ac}_N)+(1+\gamma'^2)^{-1}(J^{ac}_D\pm J_N)\nonumber\\
\tilde{J}_\pm(\vq)&=&(1+\gamma'^2)^{-1}(J^{ac}_D\pm J_N)\\
J_{D,N}&=&\frac{1}{2}(J_{D,N}^a+J_{D,N}^c)\nonumber\\
J^{ac}_{D,N}&=&\frac{1}{2}(J_{D,N}^a-J_{D,N}^c)\nonumber
\end{eqnarray}

In the case J$^a_N$= J$^c_N$ and J$^a_D$= J$^c_D$ this reduces to the
isotropic result of Eqs.(\ref{DI4},\ref{DI5}).

\subsection{Model calculation, comparison with INS results}

The simple AF structure was determined in
Ref.[\onlinecite{Krimmel92}] and subsequently inelastic neutron
scattering (INS) experiments on single crystalline \U
were performed\cite{Mason97} in an energy range up to 10 meV
in the whole AFBZ to determine the collective magnetic excitations. It
was concluded that for $\vq\parallel$ c$^*$-axis well defined magnetic
excitation modes exist. For $\vq\perp$  c$^*$-axis their dispersion
was also determined, however line widths are generally larger in this
case. An extrinsic damping mechanism via coupling to itinerant 5f-
electrons was concluded from the absence of a zone center gap. However
later more accurate low energy experiments \cite{Bernhoeft98,Sato97} have
shown that a gap of 1 meV exists at the AF wave vector (AFBZ zone
center). Within the present RPA theory the exchange anisotropy
introduced before can only generate small gaps $\leq$ 0.2 meV because
of the vicinity to the quantum critical point at $\xi_c$=1 of
the AF singlet- singlet system as discussed before. The observed zone
center gap may have a dynamical origin due to terms neglected in
RPA. In addition the $\vq\simeq$0 low energy excitation
modes will be strongly shifted and broadened due to the interaction
with 5f- conduction electrons as proposed and investigated in
Ref.(\onlinecite{Sato01}).

\begin{figure}
\centerline{\psfig{figure=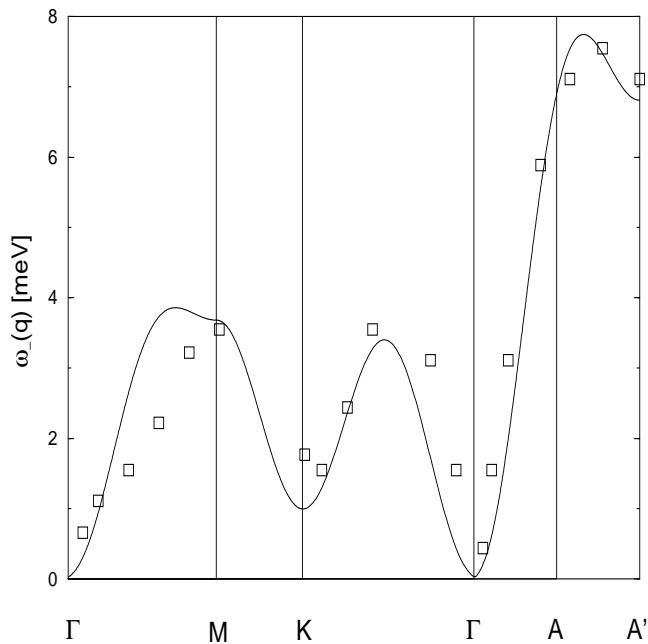,height=8.5cm,width=8.5cm,angle=-90}}
\vspace{1cm}
\caption{Magnetic exciton dispersion of UPd$_2$Al$_3$ in the
hexagonal AFBZ. Squares are data from
Ref.[9]. Solid line: Calculation from
Eq.(\ref{DIA}) using the exchange parameters J$_i$ (i=0-4) in meV
J$_0$= -1.63, J$_1$= -0.4, J$_2$= -0.31, J$_3$= 1.75 and J$_4$=
0.6 as defined in Fig. 1. For technical reasons an a-c exchange
anisotropy of 1.1 is assumed which leads to a very small $\Gamma$-
point gap.} 
\label{FIGDI2}
\end{figure}

In the present discussion the focus is on the fundamental origin of
the superconducting pair potential and the resulting gap function
within the weak coupling approach. For this purpose the details of the
low energy spectral behaviour of the soft mode (AFBZ) is not of
primary interest. It is more important to obtain a good global
description of collective magnetic excitations for all wave vectors
and energies since it is the total spectrum which determines the
strength and anisotropy of the pairing potential. For this reason we
use the RPA theory developed in the previous section to describe the
magnetic excitations found in Ref.(\onlinecite{Mason97}) for the
whole AFBZ. Going beyond RPA for the magnetic exciton spectrum would
also necessitate a strong coupling approach to the pairing problem to
stay on the same level of theoretical treatment for magnetic and
superconducting sub- systems. This challenging problem will be left
for a future investigation.

We will now discuss the experimental and theoretical dispersion of
magnetic excitons for \U unfolded along typical symmetry lines of the
hexagonal AFBZ. For the singlet- singlet system the magnetic exciton dispersion
according to Eq.(\ref{DIA}) requires the following quantities: CEF- gap
$\Delta$, exchange constants for the various neighbor shells and the
uniaxial anisotropy constant. For the CEF splitting we take $\Delta$= 6
meV which corresponds roughly to the center of the magnetic exciton
band as seen e.g. in the simple model of Sect.(III.B.2). Furthermore this is
also approximately the value obtained from fitting the susceptibility
as function of temperature \cite{Shiina01}. Since T$_N$ (14.3 K) $\ll$
$\Delta$ (66 K) we are in the CEF dominated regime: According to Eq.(\ref{CRI})
$\xi=(\tanh \frac{\Delta}{2T_N})^{-1}$= 1.015 is only slightly above
the value for the QCP at $\xi_c$=1 which leads to a saturation
moment $\la J\ra_0\ll\frac{1}{2}\alpha$ and hence
$\Delta'\simeq\Delta$. Therefore the excitation spectrum will be
almost like the magnetic excitons in the paramagnetic regime. 
The exchange model used
for the local moments of \U is illustrated in Fig.\ \ref{FIGSTR}. A
considerable  number of neighbor shells has to be included to get an
agreeable fit to the experimental data. The result is shown in Fig.\
\ref{FIGDI2} with the exchange constants in the figure caption
corresponding to the notation in Fig.\ \ref{FIGSTR}. The
calculated and experimental dispersion are strongest along the hexagonal
c-axis where the modes are also well defined i.e. the line width is
small. In addition it shows a pronounced dip at the K-point. Because
of the relation $\omega_+(\vq\pm\vQ)= \omega_-(\vq)$ only $\omega_-(\vq)$ is
plotted. In the sector $\Gamma$A it corresponds
to the acoustic mode and in the sector AA' it is the optical mode,
this is illustrated in Fig.\ \ref{FIGDI1} (left panel). The
large mode energy at A' gives direct evidence that one has a CEF- dominated
singlet- singlet system; in the opposite exchange dominated regime
with hypothetical T$_N\gg\Delta$ one would have almost degenerate
acoustic and optical branches close to $\omega_{AF}$ shown in Fig.\
\ref{FIGDI1} (right panel) and in this case the excitation energy at
A' should be very small again. It is not entirely clear how realistic
the dip at the K- point reported in Ref.(\onlinecite{Mason97}) is in
view of the fact that the line width becomes much bigger than the mode
energy at this symmetry point.

\section{The superconducting pair potential}

In the known superconductors the formation of Cooper pairs is thought
to be mediated by the exchange of bosonic excitations of the
medium. This is a simplified picture of a complicated
many body problem. More generally the pair formation is the result of
a singular behaviour in the low energy part of the two particle
scattering matrix. In the simplified picture this singularity is
described by the repeated electron-electron scattering via boson
exchange. In a nonretarded weak coupling approach this is
equivalent to the formation of a bound state in the appropriate
scattering channel. 

Two origins of Cooper pair formation are
commonly involved for describing real superconductors:
(i) the (BCS) electron-phonon mechanism in conventional superconductors
where the effective attraction needed for bound state formation is
provided by the exchange of a real frequency (propagating mode)
phonon. It is spin independent and allows only for the formation of
s- wave spin singlet Cooper pairs. 
(ii) the spin fluctuation mechanism in strongly correlated electron
systems (HF-compounds, cuprates etc.) where the Cooper pair formation is due to
exchange of an imaginary frequency spin fluctuation of conduction
electrons, predominantly close to an AF zone boundary wave vector
(overdamped "anti-paramagnon" mode). The latter mechanism which is in
principle relevant also for \U, has originally been introduced in the
context of superfluid He$^3$ by Nakajima \cite{Nakajima73} in its
paramagnon form. Later
it has been adopted for various Heavy Fermion supercoductors by many
authors, see e.g. Ref.(\onlinecite{Miyake86}) because it naturally
leads to the possibility of unconventional pair states thought to
occur in these compounds. It involves the spin degrees of freedom in a
rotationally invariant way and allows for spin singlet and triplet
pair formation.
(iii) As a novel mechanism it has recently been proposed \cite{Sato01}
that in U- compounds like \U with 5f- electrons of dual character the
exchange of magnetic excitons is at the origin of superconductivity. This
mechanism is distinctly different from those discussed before. On one
hand it is mediated by a real frequency propagating mode, the magnetic
excitons, on the other hand, as in the previous case it involves the
spin variables, but due to the presence of a CEF splitting the
rotational symmetry is broken.

In this part of the present work we will derive and discuss the
nonretarded effective pair potential of this new mechanism in 
detail. Because it is important to understand its differences 
as compared to the spin fluctuation mechanism usually
invoked for HF metals we first give a short summary of the salient
ingredients of the latter.

\subsection{The spin fluctuation model of pairing}

This pairing mechanism may be dominant in many strongly correlated
electron systems where one has a screened on-site Coulomb interaction of
heavy quasiparticles described by 

\begin{eqnarray}
\label{HIC}
H_{int}&=&I_c\int d^3rn_\da(\vec{r})n_\ua(\vec{r})\nonumber\\
&&=\frac{I_c}{V}\sum_{\vk,\vk~'}
c_{\vk\ua}^\dagger c_{-\vk\da}^\dagger c_{-\vk~'\da}c_{-\vk~'\ua}
\end{eqnarray}

Summing appropriate two particle scattering diagrams in RPA
\cite{Nakajima73} Eq.(\ref{HIC}) leads to an effective pair
Hamiltonian ($\vq=\vk~'-\vk$)

\begin{eqnarray}
\label{HSF}
H_{eff}&=&\frac{1}{2}\sum_{\vk\vk',\alpha\beta\gamma\delta}
[V_\rho(\vq)\delta_{\alpha\beta}\delta_{\gamma\delta}
+V_s(\vq)\vec{\sigma}_{\alpha\beta}\vec{\sigma}~'_{\gamma\delta}]\cdot
\nonumber\\
&&c_{\vk\alpha}^\dagger c_{-\vk\gamma}^\dagger
c_{-\vk'\delta}c_{\vk'\beta}
\end{eqnarray}

where the sum in parentheses describes the effective two particle
interaction and $\vec{\sigma},\vec{\sigma}~'$ denote Pauli matrices.
It contains a spin- independent potential scattering of strength
V$_\rho(\vq)$  and an exchange scattering of strength V$_s(\vq)$ which
are obtained in RPA as 

\begin{equation}
\label{POT}
V_\rho(\vq)=\frac{\frac{1}{2}I_c}{1+I_c\chi_0(\vq)};\;\;\;\;\;\;
V_s(\vq)=\frac{-\frac{1}{2}I_c}{1-I_c\chi_0(\vq)}
\end{equation}

Using the projectors for pair spin singlet (S=0) and triplet (S=1)
channels respectively,

\begin{eqnarray}
\label{PR1}
P_0&=&\frac{1}{4}(1-\vec{\sigma}\vec{\sigma}~')\nonumber\\
P_1&=&\frac{1}{4}(3+\vec{\sigma}\vec{\sigma}~')
\end{eqnarray}

with the Kronecker product
$(\vec{\sigma}\vec{\sigma}~')_{\alpha\beta;\gamma\delta}$= 
$\vec{\sigma}_{\alpha\beta}\vec{\sigma}~'_{\gamma\delta}$, one can
separate V$_{eff}(\vq)$ into the irreducible potentials of singlet and
triplet Cooper pairing: 

\begin{eqnarray}
\label{V01}
V_0(\vq)&=&V_\rho(\vq)-3V_s(\vq)\simeq \frac{3}{2}I_c^2\chi(\vq)\nonumber\\
V_1(\vq)&=&V_\rho(\vq)+ V_s(\vq)\simeq -\frac{1}{2}I_c^2\chi(\vq)
\end{eqnarray}

where the last approximation in the above equations holds for wave
vectors with enhanced spin fluctuations, i.e. when
$\chi(\vq)\gg\chi_0(\vq)$. Here

\begin{equation}
\label{SUS}
\chi(\vq)=\frac{\chi_0(\vq)}{1-I_c\chi_0(\vq)}
\end{equation}

is the RPA susceptibility of conduction electrons which has its
maximum at a nesting vector $\vQ$ which may be the commensurate AF
vector. The effective potential may be written,
using spin projectors P$_0$, P$_1$ and defining the sequence of states
as ($\ua\ua,\ua\da,\da\ua,\da\da$), in the following form:  

\begin{eqnarray}
\label{PMA}
&&\tensor{V}_{eff}(\vq)=V_0(\vq)P_0+V_1(\vq)P_1=\nonumber\\
&&\left(\matrix{
V_1& 0 & 0 & 0 &\cr
0 &  \frac{1}{2}(V_0+V_1)& -\frac{1}{2}(V_0-V_1)  & 0 &\cr
0 & -\frac{1}{2}(V_0-V_1)&  \frac{1}{2}(V_0+V_1)  & 0 &\cr
0 & 0 & 0 & V_1&\cr
}\right)
\end{eqnarray}

where according to Eq.(\ref{V01})
$\frac{1}{2}(V_0+V_1)\simeq\frac{1}{2}I_c^2\chi,
-\frac{1}{2}(V_0-V_1)\simeq -\frac{1}{2}I_c^2\chi, V_1\simeq
-\frac{1}{2}I_c^2\chi$. Note that the total spin component
s$_z$=$\frac{1}{2}$($\sigma_z+\sigma'_z$) is conserved due to rotational
invariance of V$_{eff}$, therefore matrix elements corresponding to a
change of the z- component of the pair spin like
$\la\ua\ua|V_{eff}|\da\da\ra$ = $\la\da\da|V_{eff}|\ua\ua\ra$
vanish. One may explicitly check by
diagonalisation that V$_{eff}(\vq)$ has singlet $|0\ra$ and triplet
$|1,s_z=0,\pm1\ra$ eigenstates with energies V$_0$ and V$_1$
respectively. The momentum dependence of $\chi(\vq)$ finally
determines which orbital
pair state will be realized as disussed in Ref.[\onlinecite{Miyake86}]
who conclude that even parity singlet pairs are favored in general.

\subsection{Pair potential due to magnetic exciton exchange}

In the dual model of Eq.(\ref{HAM}) the on-site Coulomb interaction of the
itinerant 5f- electrons is neglected and only indirect interaction via
intermediate CEF excitations due to the last term is
considered. Since the latter can propagate this means an effective
conduction electron interaction via the exchange of magnetic
excitons. This will now be derived by similar diagrammatic methods as
in Ref.[\onlinecite{Nakajima73}], however the details will be quite
different for the present case. In this section we neglect the main
effect of AF order on the pair potential and consider only the
paramagnetic case. For the derivation of the pair potential the Hamiltonian
Eq.(\ref{HAM}) will be first rewritten in terms of more convenient bosonic
variables for the CEF singlet-singlet excitations introduced via the
pseudo spin Holstein- Primakoff representation

\begin{equation}
\label{HOL} 
S_i^+=a_i\;\; S_i^-=a^+_i\;\; S_i^z= \frac{1}{2}-a_i^+a_i 
\end{equation}

Here a$_i$ is a local boson that describes CF- excitations
$|g\ra\rightarrow |e\ra$ at site i. For T$\ll\Delta$ the influence of
unphysical states $(a_i)^n|0\ra$ with n$\geq$2 is
negligible. This condition is very well fulfilled since T$\ll T_N\ll\Delta$. 
Then the on- site exchange

\begin{equation}
\label{CF1}
H_{cf}= -2I_0(g-1)\sum_i[s_{iz}J_{iz}+
s_i^+J_i^-+s_i^-J_i^+]
\end{equation}

can be transformed, using J$_z=\epsilon (\frac{1}{2}-S_z$) and J$^\pm
=\alpha S^\pm$ and Eq.(\ref{HAM}) to obtain

\begin{eqnarray}
\label{CF2}
H_{cf}&=&-\frac{I}{\sqrt{N}}\sum_{\vk\vk'}
[c^+_{\vk'\ua}c_{\vk\da}a^+_{\vk-\vk'}+c^+_{\vk'\da}c_{\vk\ua}a_{\vk'-\vk}]
\nonumber\\
&&-\frac{I_z}{\sqrt{N}}\sum_{\vk\vk'\vq}
(c^+_{\vk'\ua}c_{\vk\ua}-c^+_{\vk'\da}c_{\vk\da})a^+_{\vq+\vk-\vk'}a_{\vq}
\end{eqnarray}

where I=$\alpha$(g-1)I$_0$ and I$_z$=$\epsilon$(g-1)I$_0$. For
$\alpha, \epsilon$ see Eq.(\ref{JOP}), furthermore

\begin{equation}
\label{BOS}
a_{\vq}=\frac{1}{\sqrt{N}}\sum_ia_ie^{i\vq\vec{R}_i}
\end{equation}

creates CEF- bosons of momentum $\vq$. The total Hamiltonian of Eq.(\ref{HAM})
is then given by 

\begin{equation}
\label{HBO}
H=\sum _{\vk\sigma}\epsilon_{\vk\sigma}c^\dagger_{\vk\sigma}c_{\vk\sigma}
+ \Delta\sum_{\vq}a^+_{\vq}a_{\vq} +H_{ff} +H_{cf}
\end{equation}

\begin{figure}
\centerline{\psfig{figure=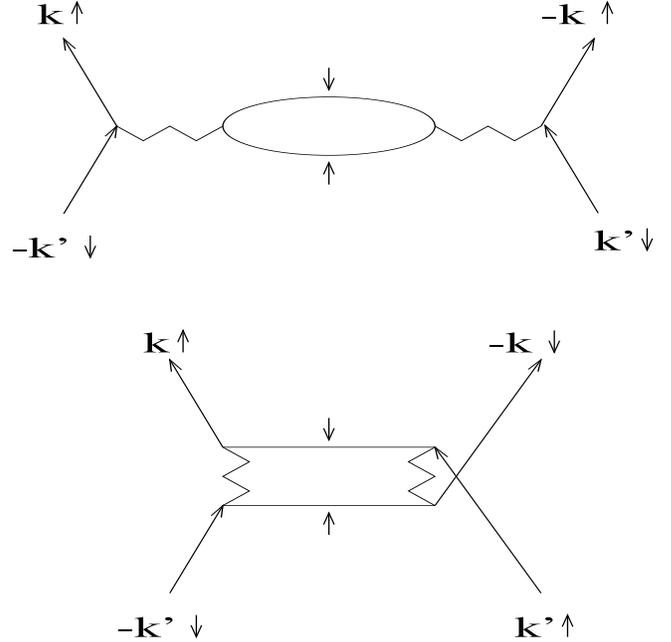,height=8.5cm,width=8.5cm}}
\vspace{1cm}
\caption{Lowest order ($\sim$I$^4$) polarisation diagram (above) with
$\vq=\vk'+\vk$ and
exchange diagram (below) with $\vq=\vk'-\vk$ contributing to the
scattering potentials V$_a$ and V$_d$ of Eq.(\ref{POE}) respectively. The
zig-zag line represents the boson propagator D$_0$ and the bubble is
equal to the conduction electron
susceptibility $\chi_0(\vq)$ in the nonretarded approximation. Small
arrows represent conduction electrons spin. Note
that the upper diagram describes a process which does not conserve the
total pair spin component (s$_z$+s'$_z$) leading to a pair potential
Eq.(\ref{FIN}) which is not rotationally invariant in spin space. Summation of
diagrams of odd order in I$^2$ leads to V$_b$ and V$_c$.} 
\label{FIGDIA}
\end{figure}

The interaction term in Eq.(\ref{CF2}) contains both 1- boson absorption
or creation parts as well as 2- boson scattering parts. The
latter describes conduction electron scattering from quantum
fluctuations in the occupation of the two singlet CEF- states. Its
contribution will be neglected in the following discussion.
The boson propagator is given by

\begin{equation}
\label{DBO}
D_{\vq}(\tau)=D^+_{\vq}(-\tau)= -\la T_\tau
a_{\vq}(\tau)a_{\vq}^+(0)\ra
\end{equation}

and its Fourier transform ($\omega_n=2n\pi$T) by

\begin{equation}
\label{GBO}
D_{\vq}(i\omega_n)=D^+_{\vq}(-i\omega_n)
=\frac{1}{i\omega_n-\Delta}
\end{equation}

The momentum $\vq$ for the noninteracting CEF- bosons is a dummy
variable. Note that in the zero frequency limit
(D$_{\vq}$D$^+_{\vq}$)$_{\omega_n\rightarrow 0}$=$\frac{1}{\Delta^2}$
The lowest order nontrivial polarisation and exchange diagrams for the
effective e-e interaction vertex due to the 1- boson part of
Eq.(\ref{CF2}) are of order $\sim$I$^4$ and are shown in Fig.\
\ref{FIGDIA}. For the nonretarded e-e
vertex function $\Gamma_0(\vq)$ that enters the effective pairing
Hamiltonian we only need the zero frequency limit of Fig.\ \ref{FIGDIA} in
which the diagrams lead to equal contributions:

\begin{equation}
\label{VER}
\Gamma_0^2(\vq)=
[I^2\chi_0(\vq,0)]^2(D_{\vq}D^+_{\vq})_{\omega_n\rightarrow 0}=
[\frac{I^2}{\Delta}\chi_0(\vq,0)]^2
\end{equation}

Here the noninteracting susceptiblility
$\chi_0(\vq)=\chi_{xx}(\vq)=\chi_{yy}(\vq)$ is given by the Lindhard function 

\begin{equation}
\label{LIN}
\chi_0(\vq)=\sum_{\vk}\frac{f_{\vk}-f_{\vk+\vq}}
{\epsilon_{\vk+\vq}-\epsilon_{\vk}}
\end{equation}

The summation of polarisation and exchange diagrams as in Fig.\
\ref{FIGDIA} to infinite order
in RPA leads to the following nonretarded e-e
interaction potential V$_{\alpha\beta\gamma\delta}(\vq)$ for the
various channels of initial ($\beta\delta$) and final ($\alpha\gamma$)
spin configurations:

\begin{eqnarray}
\label{POE}
V_a&=&V^a_{\da\da,\ua\ua}=V^a_{\ua\ua,\da\da}=\nonumber\\
&&-\frac{\chi_0^{-1}\Gamma_0^2}{1-\Gamma_0^2}=
-\frac{1}{2}I^2[\frac{1}{\Delta(1-\Gamma_0)}-\frac{1}{\Delta(1+\Gamma_0)}]
\nonumber\\
V_b&=&V^b_{\da\ua,\ua\da}=V^a_{\ua\da,\da\ua}=\\
&&\frac{I^2D_0}{1-\Gamma_0^2}=
-\frac{1}{2}I^2[\frac{1}{\Delta(1-\Gamma_0)}+\frac{1}{\Delta(1+\Gamma_0)}]
\nonumber\\
V_c&=&V^c_{\da\da,\ua\ua}=V^c_{\ua\ua,\da\da}=V_b \nonumber\\
V_d&=&V^d_{\da\ua,\ua\da}=V^c_{\ua\da,\da\ua}=V_a \nonumber
\end{eqnarray} 

Here V$_a(\vq)$, V$_b(\vq)$ with momentum transfer $\vq=\vk+\vk'$
correspond to the polarisation and V$_c(\vq)$, V$_d(\vq)$ with 
$\vq=\vk-\vk'$ to the exchange diagrams of Fig.\ \ref{FIGDIA}. Adding up all
contributions and transforming $\vk\rightarrow -\vk$ in the first two
(a,b) terms one obtains the effective e-e interaction Hamiltoninan
where now $\vq=\vk-\vk'$:

\begin{eqnarray}
\label{EXC}
&&H_{eff}=\nonumber\\
&&\sum_{\vk\vk'}
[V_a(\vq)+V_c(\vq)]
[c_{\vk\ua}^\dagger c_{-\vk\ua}^\dagger c_{-\vk'\da}c_{\vk'\da}+
c_{\vk\da}^\dagger c_{-\vk\da}^\dagger
c_{-\vk'\ua}c_{\vk'\ua}]\nonumber\\
&&+\sum_{\vk\vk'}
V_b(\vq)
[c_{\vk\da}^\dagger c_{-\vk\ua}^\dagger c_{-\vk'\da}c_{\vk'\ua}+
c_{\vk\ua}^\dagger c_{-\vk\da}^\dagger c_{-\vk'\ua}c_{\vk'\da}]\\
&&+\sum_{\vk\vk'}
V_d(\vq)
[c_{\vk\ua}^\dagger c_{-\vk\da}^\dagger c_{-\vk'\da}c_{\vk'\ua}+
c_{\vk\da}^\dagger c_{-\vk\ua}^\dagger
c_{-\vk'\ua}c_{\vk'\da}] \nonumber
\end{eqnarray}

Defining the effective e-e interaction potential by

\begin{equation}
\label{HXC}
H_{eff}=\frac{1}{2}\sum_{\vk\vk',\alpha\beta\gamma\delta}
V^{eff}_{\alpha\beta\gamma\delta}(\vq)
c_{\vk\alpha}^\dagger c_{-\vk\gamma}^\dagger
c_{-\vk'\delta}c_{\vk'\beta}
\end{equation}

we get the explicit effective interaction matrix by comparison with
the previous equation, using V$_\pm$=V$_a\pm$V$_b$ and again defining
the state sequence as ($\ua\ua,\ua\da,\da\ua,\da\da$):

\begin{eqnarray}
\label{PXC}
&&\frac{1}{2}\tensor{V}_{eff}(\vq)=\nonumber\\
&&\left(\matrix{
0   &  0                    & 0                      & V_+ &\cr
0   &  \frac{1}{2}(V_-+V_+) &  -\frac{1}{2}(V_--V_+) & 0   &\cr
0   & -\frac{1}{2}(V_--V_+) &   \frac{1}{2}(V_-+V_+) & 0   &\cr
V_+ &  0                    & 0                      & 0   &\cr
}\right)
\end{eqnarray}

Comparison with Eq.(\ref{PMA}) shows that interaction components
V$^{eff}_{\ua\ua\ua\ua}$ etc. are absent due to the fact that boson
absorption or emission always leads to spin
flips. On the other hand there are additional components like
V$^{eff}_{\ua\ua\da\da}$ =V$_a$+V$_b$=V$_+$ which do not conserve the total
pair spin component s$^t_z$=(s$_z$+s'$_z$). Thus magnetic exciton
exchange does {\em not} lead to a rotationally invariant effective
pairing potential as in the case of the spin fluctuation mechanism in
Sect. (IV.A). This is a direct consequence of the presence of a CEF
splitting. The decomposition of V$_{eff}$ into irreducible
contributions as previously given in Eqs.(\ref{PMA}) is therefore more
involved and will be discussed below.

The static e-e vertex function $\Gamma'_0(\vq)$=
$\Gamma'_0(\vq)$-N$^{-1}\sum_{\vq}\Gamma_0(\vq)$ is related to the
Fourier transform of the effective exchange interaction
J$_{RKKY}(\vq)$ of pseudo spins via the the relations

\begin{eqnarray}
\label{RKK}
\alpha^2\la S\ra J_{RKKY}(\vq)&=&I^2\chi'_0(\vq)\nonumber\\
\Gamma'_0(\vq)=\frac{I^2}{\Delta}\chi'_0(\vq)&=&
\frac{\alpha^2\la S\ra}{\Delta}J_{RKKY}(\vq)
\end{eqnarray}

where $\chi'_0(\vq)$ was given in Eq.(\ref{EFF},\ref{LIN}). In
addition the pseudo spins
have a superexchange J$_{ff}(\vq)$ whose contribution in static RPA
limit is described by equivalent diagrams thus leading to a total
vertex function

\begin{eqnarray}
\label{JTO}
\Gamma(\vq)=\Gamma'_0(\vq)+\Gamma_{ff}(\vq)=\nonumber\\
\frac{\alpha^2\la S\ra}{\Delta}[J_{RKKY}(\vq)+J_{ff}(\vq)]&\equiv&
\frac{\alpha^2\la S\ra}{\Delta}J(\vq)
\end{eqnarray}

This result may now be substituted in Eq.(\ref{POE}). As mentioned in
Sect.(II) the total exchange function J$(\vq)$ will be parametrized by
comparison with the experimental exciton dispersion. It has intra- and
inter- sublattice parts J$_D(\vq)$ and J$_N(\vq)$ respectively with
J$(\vq)$=J$_D(\vq)$+J$_N(\vq)$. Absorbing the constant
$\frac{1}{N}\sum_{\vq}\Gamma_0(\vq)$ into a renormalized CEF splitting
$\Delta$ and replacing  $\Delta\rightarrow\Delta'$ for the AF ordered
case we then obtain, defining $\omega_R(\vq)=2\Delta'-\omega_-(\vq)$ 

\begin{eqnarray}
\label{VXC}
\Delta'[1-\Gamma(\vq)]&=&
\Delta'-\alpha^2\la S\ra J_D(\vq)-\alpha^2\la S\ra J_N(\vq)\equiv
\omega_-(\vq)\nonumber\\
&&\\
\Delta'[1+\Gamma(\vq)]&=&
\Delta'+\alpha^2\la S\ra J_D(\vq)+\alpha^2\la S\ra J_N(\vq)\equiv
\omega_R(\vq)\nonumber
\end{eqnarray}

Here $\omega_-(\vq)$ is nothing but the excitonic dispersion branch
which becomes soft at the zone boundary $\vQ$ of the PMBZ at T$_N$ and
evolves into the acoustic branch with its minimum at the new $\Gamma$-
point of the AFBZ as explained in Sect.(III) Note that the energy 
$\omega_R(\vq)$=$2\Delta'-\omega_-(\vq)$ does {\em not} correspond to
an excitation branch. Since 2$\Delta'\gg\omega_-(\vq)$ for all
realistic cases $\omega_R(\vq)$ will never be small and
$\omega_R(\vq)^{-1}$ is regular throughout the AFBZ. 
Using Eq.(\ref{VXC}) the effective potentials of Eq.(\ref{VIR}) can
now be written as

\begin{eqnarray}
\label{VAB}
V_a&=&-\frac{I^2}{2}[\frac{1}{\omega_-(\vq)}-\frac{1}{\omega_R(\vq)}]
\nonumber\\
V_b&=&-\frac{I^2}{2}[\frac{1}{\omega_-(\vq)}+\frac{1}{\omega_R(\vq)}]
\end{eqnarray}

Inserting this result into the interaction matrix of Eq.(\ref{PXC}) the
possible irreducible pair potentials (i.e. the eigenvalues V$_\ka$,
$\ka$=0,u,v,w of $\tensor{V}_{eff}$) are given by

\begin{eqnarray}
\label{VIR}
V_0&=& (V_a-V_b)= I^2\frac{1}{\omega_R(\vq)}\nonumber\\
V_u&=&-(V_a+V_b)= I^2\frac{1}{\omega_-(\vq)}\nonumber\\
V_v&=& (V_a+V_b)=-I^2\frac{1}{\omega_-(\vq)}\\
V_w&=& (V_a+V_b)=-I^2\frac{1}{\omega_-(\vq)}\nonumber
\end{eqnarray}

with the correspondence V$_0$=V$_-$, V$_u$= -V$_v$= -V$_w$= V$_+$ to
Eq.(\ref{PXC}). The pair states obtained as eigenstates from Eq.(\ref{PXC})
are classified according to their s$_z$=$\uparrow\downarrow$
(=$\pm\frac{1}{2}$) quantum numbers. Later on when the effect of AF
order is considered, a representation with s$_x$=+,-
(=$\pm\frac{1}{2}$) may be useful. It is obtained by
$|+\ra=R_y(\frac{\pi}{2})|\ua\ra$ etc. where
$R_y=\frac{1}{\sqrt{2}}(1-i\sigma_y)$ is a rotation by $\frac{\pi}{2}$
around the y- axis. Therefore we also give the relation between the
two notations for the eigenstates $\psi_{\ka}$ or $\psi_{\ka}^x$ which
correspond to V$_{\ka}, \ka$=0,u,v,w: 

\begin{eqnarray}
\label{PSI}
\psi_0=\psi_0^x&=&\frac{1}{\sqrt{2}}(|\ua\da\ra -|\da\ua\ra)\nonumber\\
\psi_u=-\psi_w^x&=&\frac{1}{\sqrt{2}}(|\ua\ua\ra -|\da\da\ra)\nonumber\\
\psi_v=\psi_v^x&=&\frac{1}{\sqrt{2}}(|\ua\ua\ra +|\da\da\ra)\\
\psi_w=\psi_u^x&=&\frac{1}{\sqrt{2}}(|\ua\da\ra +|\da\ua\ra)\nonumber
\end{eqnarray}

Here $\psi_0$ describes singlet pairs. In the spin fluctuation
model ($\psi_{\ka}, \ka$= u,v,w) correspond to the degenerate triplet
pairs. In the present magnetic exciton model according to Eq.(\ref{VIR})
obviously this degeneracy is partly lifted because the basic
Hamiltonian Eq.(\ref{CF2},\ref{HBO}) is not rotationally invariant in
spin space as a result of the coupling to CEF split localized
states. The triplet is split into a nondegenerate state $\psi_u$ and a doublet
($\psi_v$,$\psi_w$). 
As a consequence in the present case the irreducible projector
representation for the pair potential like Eq.(\ref{PXC}) is different. The
projectors to the eigenvectors which satisfy P$_i$P$_j$
=P$_i\delta_{ij}$ may be written as

\begin{eqnarray}
\label{PR2}
P_0=P_0^x&=&\frac{1}{4}
(1-\vec{\sigma}\vec{\sigma}')\nonumber\\
P_u=P_w^x&=&\frac{1}{4}
(1-\sigma_x\sigma_x'+\sigma_y\sigma_y'+\sigma_z\sigma_z')\nonumber\\
P_v=P_v^x&=&\frac{1}{4}
(1+\sigma_x\sigma_x'-\sigma_y\sigma_y'+\sigma_z\sigma_z')\\
P_w=P_u^x&=&\frac{1}{4}
(1+\sigma_x\sigma_x'+\sigma_y\sigma_y'-\sigma_z\sigma_z')\nonumber
\end{eqnarray}

The projector to the doublet subspace P$_D$=(P$_v$+P$_w$) and to
the total triplet subspace P$_1$=(P$_u$+P$_v$+P$_w$) are given by 

\begin{eqnarray}
\label{PR3}
P_D&=&\frac{1}{2}(1+\sigma_x\sigma_x')\nonumber\\
P_1&=&\frac{1}{4}(3+\vec{\sigma}\vec{\sigma}')
\end{eqnarray}

Then similar as in Eq.(\ref{PMA}) V$_{eff}(\vq)$ as mediated by magnetic
excitons (Eq.(\ref{PXC})) may finally be written as 

\begin{equation}
\label{FIN}
\tensor{V}_{eff}(\vq)=
\frac{I^2}{\omega_R(\vq)}P_0 +\frac{I^2}{\omega_-(\vq)}P_u
-\frac{I^2}{\omega_-(\vq)}(P_v+P_w)
\end{equation}

There is an intuitively simple interpretation for the CEF- induced
triplet splitting present in this model: The conduction electrons can
scatter from virtual CEF- excitations $\Delta$ only with their s$_x$
(or s$_y$)- component, therefore the pair states will experience a
potential $\sim(I^2/\Delta)$(s$_x^{tot})^2$ in lowest order which splits
the ($\psi^x_{\ka}, \ka$=u,v,w) triplet into nondegenerate
$\psi^x_w$ (s$_x^{tot}$=0) and $\psi^x_u\pm\psi^x_v$
(s$_x^{tot}=\pm$1) degenerate pair states.  

The previously developed theory of magnetic excitons which explains
the experimental observations as shown in Fig.\ \ref{FIGDI2} can now be used to
calculate the pair potential from Eqs.(\ref{VIR},\ref{FIN}). However
only the q$_x$, q$_y$- averaged potential is needed for the gap
equations as explained in the next section 

\section{The superconducting gap equations}

The effective e-e interaction may lead to a
superconducting state with an order parameter given by the gap matrix
($\vq=\vk-\vk'$) 

\begin{equation}
\label{GA1}
\Delta_{\alpha\beta}^{\vk}= 
-\sum_{\vk'\gamma\delta}V^{\vk-\vk'}_{(\alpha\beta),(\gamma\delta)}
\la c_{\vk'\gamma}c_{-\vk'\delta}\ra
\end{equation}

which satisfies the m.f. gap equation

\begin{equation}
\label{GA2}
\Delta_{\alpha\beta}^{\vk}=
-\sum_{\vk'\gamma\delta}V^{\vk-\vk'}_{(\alpha\beta)(\gamma\delta)}
\frac{\Delta_{\gamma\delta}^{\vk'}}{2E_{\vk'}}
\tanh\frac{1}{2}\beta E_{\vk'}
\end{equation}

Here we defined V$^{\vk-\vk'}_{(\alpha\beta)(\gamma\delta)}$= 
V$^{\vk-\vk'}_{\alpha\gamma\beta\delta}$ and 
$(\alpha\beta), (\gamma\delta)$ = $\ua\ua, \ua\da, \da\ua,
\da\da$  are double spin indices in the notation for s$_z$-
quantisation and E$_{\vk}$ are the superconducting quasiparticle
energies. The gap matrix may be decomposed in the canonical way as 

\begin{equation}
\label{GA3}
\tensor{\Delta}^{\vk}=i\sum_{i=0}^3\sigma_i\sigma_2 d_i^{\vk}
\end{equation}

where d$_0^{\vk}$, $\vec{d}^{\vk}$=(d$_u^{\vk}$,d$_v^{\vk}$,d$_w^{\vk}$)
correspond to the pair states $\psi_0$, ($\psi_u$,$\psi_u$,$\psi_v$)
respectively introduced before. Explicitly we have the relations in both s$_z$-
and s$_x$ quantisation notation:

\begin{eqnarray}
\label{DVC}
d_0^{\vk}&=&\frac{1}{2}(\Delta^{\vk}_{\ua\da}-\Delta^{\vk}_{\da\ua})
           =\frac{1}{2}(\Delta^{\vk}_{+-}-\Delta^{\vk}_{-+})\nonumber\\
d_u^{\vk}&=&-\frac{1}{2}(\Delta^{\vk}_{\ua\ua}-\Delta^{\vk}_{\da\da})
           =\frac{1}{2}(\Delta^{\vk}_{+-}+\Delta^{\vk}_{-+})\nonumber\\
d_v^{\vk}&=&\frac{1}{2i}(\Delta^{\vk}_{\ua\ua}+\Delta^{\vk}_{\da\da})
           =\frac{1}{2i}(\Delta^{\vk}_{++}+\Delta^{\vk}_{--})\\
d_w^{\vk}&=&\frac{1}{2}(\Delta^{\vk}_{\ua\da}+\Delta^{\vk}_{\da\ua})
           =-\frac{1}{2}(\Delta^{\vk}_{++}-\Delta^{\vk}_{--})\nonumber
\end{eqnarray}

Using the projector representation Eq.(\ref{FIN}) for the effective
potential and the identity

\begin{eqnarray}
\label{PGA}
P_0\vec{\Delta}_{\vk}&=&d_0^{\vk}\nonumber\\
P_{\ka}\vec{\Delta}_{\vk}&=&d_{\ka}^{\vk}\hat{d}_{\ka}\;\;(\ka\neq 0)
\end{eqnarray}

where $\vec{\Delta}^{\vk}$=\{$\Delta_{(\alpha\beta)}$\} is a four
component vector acted on by the 4$\times$4 projector matrices of
Eq.(\ref{PR2}). We then obtain the scalar irreducible gap equations ($\ka$=0,u,v,w)

\begin{eqnarray}
\label{DGA}
d_{\ka}^{\vk}&=&-\sum_{\vk'}V_{\ka}^{\vk-\vk'}F_{\vk'}d_{\ka}^{\vk'}
\nonumber\\
F_{\vk}&=&\frac{1}{2}E_{\vk}\tanh\frac{1}{2}\beta E_{\vk}\\
E_{\vk}&=&[(\epsilon_{\vk}-\mu)^2+|d_{\ka}^{\vk}|^2]^{\frac{1}{2}}
\nonumber
\end{eqnarray}

Before solving these equations explicitly we note that the 
gap functions can be classified according to the crystal symmetry group which
in this case is the orthorhombic group D$_{2h}$ due to the presence of
the AF order parameter with $\vec{m}_{\vQ}\parallel\hat{x}$. The 
corresponding irreducible representations of gap functions can be
listed as 

\begin{eqnarray}
\label{DSY}
d_0^{\vk}&&\sim\cos k_z:\;\;A_{1g}(\Gamma_1^+)\nonumber\\
\vec{d}_{\vk}=(d_u^{\vk},0,0)&&\sim\hat{x}\sin k_z:\;\;
B_{2u}(\Gamma_2^-)\nonumber\\
\vec{d}_{\vk}=(0,d_v^{\vk},0)&&\sim\hat{y}\sin k_z:\;\;
B_{3u}(\Gamma_4^-)\\
\vec{d}_{\vk}=(0,0,d_w^{\vk})&&\sim\hat{z}\sin k_z:\;\;
A_{1u}(\Gamma_1^-)\nonumber
\end{eqnarray}

These representations have even (d$_0^{\vk}$) and odd
($\vec{d}^{~\vk}_{\ka}, \ka=u,v,w$) parity respectively The basis
functions have all odd character with respect to the transformation
$\vk\rightarrow\vk\pm\vQ$, namely
$\vec{d}_{\ka}^{\vk\pm\vQ}$= -$\vec{d}_{\ka}^{\vk}$ ($\ka$= 0,u,v,w). In
D$_{2h}$ there is no symmetry reason why (d$_v^{\vk}$,d$_w^{\vk}$) should be
degenerate. In fact we show later that the AF order connected with
D$_{2h}$ symmetry will lead to a splitting into nondegenerate
d$_-^{\vk}$ and d$_+^{\vk}$ superconducting states.

\subsection{Solution of gap equations for cylindrical symmetry}

The explicit solution of Eq.(\ref{DGA}) using the potentials of
Eq.(\ref{VIR}) is greatly facilitated by taking into account the
corrugated cylindrical FS shape modeled by Eq.(\ref{DIS}) and shown
schematically in the inset of Fig.\ \ref{FIGOPA}. In this case
Eq.(\ref{DGA}) can be approximated by a one- dimensional integral
equation in the variable k$_z$. Explicitly then

\begin{equation}
\label{DG1}
d_{\ka}(k_z)=-\sum_{k'_z}\bar{V}_i(k_z-k'_z)d_{\ka}(k'_z)
\sum_{k'_\perp}\frac{\tanh\frac{\beta}{2}E(k'_\perp,k'_z)}
{E(k'_\perp,k'_z)}
\end{equation}

where $\vk_{\perp}$=(k$_x$,k$_y$), k$_x$=k$_\perp\cos\phi$,
k$_y$=k$_\perp\sin\phi$ and $\phi$ is the azimuthal angle in the ab-
plane. For the above separation of variables it is assumed that the
$\phi$- dependence of the order parameter can be neglected, this means
we restrict to representations of the type given in Eq.(\ref{DSY}) and higher
harmonics. In addition for the cylindrical FS sheet in \U
$\epsilon(\vk_{\perp},\sigma)$ depends only on the modulus
k$_\perp$. Therefore only the q$_x$,q$_y$- {\em averaged} pair
potentials $\bar{V}_{\ka}$(q$_z$) appear in the gap equation where we
define 

\begin{equation}
\label{VAV}
\bar{V}_{\ka}(q_z)=\frac{1}{4\pi^2}\int_{BZ}dq_xdq_y
V_{\ka}(q_x,q_y,q_z)
\end{equation}

\begin{figure}
\centerline{\psfig{figure=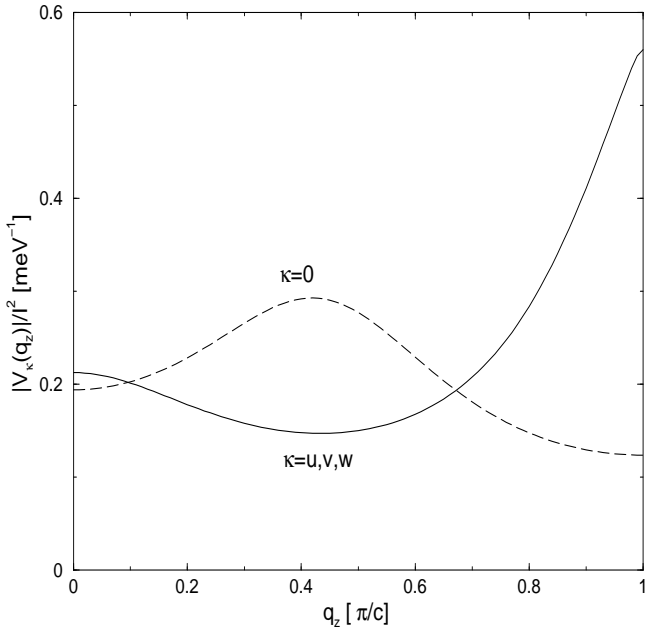,height=8.5cm,width=8.5cm,angle=-90}}
\vspace{1cm}
\caption{q$_z$- dependence of the absolute valued of q$_x$, q$_y$- averaged 
pairing potentials $\bar{V}_{\kappa}$ in the PMBZ for even
($\bar{V}_0>0$) and odd ($\bar{V}_u>0$ and $\bar{V}_v$, $\bar{V}_w
<0$) parity of the gap function. q$_z$=0 corresponds to the  A'- point and
q$_z$=$\frac{\pi}{c}$ to the $\Gamma$ point of the AFBZ in Fig. 4
respectively.}
\label{FIGPOZ}
\end{figure}

Using the expressions in Eq.(\ref{VIR}) and the calculated exciton dispersion
for \U in Sect.(III) we obtain these averaged pair potentials as shown in
Fig.\ \ref{FIGPOZ}. Because of the singular behaviour of
V$_{\ka}(\vq)$ at the AF- point $\vQ$ the absolute
value of averaged potentials $|\bar{V}_{\ka}(q_z)|$ ($\ka$=u,v,w) show
a pronounced maximum at $\vQ$=(0,0,$\pi$). On the other hand the
averaged singlet pair potential $\bar{V}_0$(q$_z$) is rather smooth
with a flat maximum for q$_z$ around 0.4$\pi$. This is due to the
nonsingular nature of  V$_0(\vq)$ as discussed in Sect.(IV).

The gap equation Eq.(\ref{DG1}) may now be completely factorized by expanding 
$\bar{V}_{\ka}$(q$_z$) (q$_z$=k$_z$-k'$_z$) in lattice harmonics according to 

\begin{eqnarray}
\label{FOU}
\bar{V}_{\ka}(q_z)&=&\sum_m\bar{V}_{\ka}^m\cos q_z=\nonumber\\
&&\sum_m\bar{V}^m_{\ka}(\cos mk_z\cos mk'_z-\sin mk_z\sin mk'_z)
\end{eqnarray}

Furthermore we may expand the even ($\ka$= 0) and odd parity
($\ka$= u,v,w) gap functions into

\begin{eqnarray}
\label{DFO}
d_0(k_z)=\sum_{m\geq 0}d^m_0\cos mk_z;\;\;\;
d_{\ka}(k_z)=\sum_{m\geq 1}d^m_{\ka}\sin mk_z
\end{eqnarray}

The order parameter may now be in general represented as vectors
$\vec{d}_0=\{d_0^m\}$  and $\vec{d}_{\ka}=\{d_{\ka}^m\}$ and the gap
equations for the components obtained from inserting
Eqs.(\ref{FOU},\ref{DFO}) in Eq.(\ref{DG1}) finally can be written as

\begin{eqnarray}
\label{DG2}
1&=&\sigma_{\ka}N(0)\bar{V}_{\ka}^mG_m[\vec{d}_{\ka}]\nonumber\\
G_m[\vec{d}_{\ka}]&=&\frac{1}{\pi}\int_0^{\pi}
\alpha_m(k'_z)F[T,k'_z,\vec{d}_{\ka}]dk'_z\\
F[T,\vec{d}_{\ka}(k'_z)]&=&\int_0^{\omega_-^c}d\xi
\frac{\tanh\frac{1}{2}\beta E(\xi,k'_z)}{E(\xi,k'_z)}\nonumber
\end{eqnarray}

Here $\omega_-^c$ is the magnetic exciton band cutoff energy,
$\sigma_{\ka}$=-1 for $\ka=0$ and $\sigma_{\ka}$=1 for
$\ka=u,v,w$. Note that $\sigma_{\ka}$ is due to the different signs for
even and odd parity parts in the harmonic decomposition in Eq.(\ref{FOU}).
Furthermore  we have for $\ka$=0:

\begin{eqnarray}
\label{QP1}
E(\xi,k_z)&=&
[\xi^2+(\sum_nd_{\ka}^n\cos nk_z)^2]^{\frac{1}{2}}\nonumber\\ 
\alpha_m(k_z)&=&\cos^2(mk_z)
\end{eqnarray}

and for $\ka$= u,v,w similar equations

\begin{eqnarray}
\label{QP2}
E(\xi,k_z)&=&
[\xi^2+(\sum_nd_{\ka}^n\sin nk_z)^2]^{\frac{1}{2}}\nonumber\\ 
\alpha_m(k_z)&=&\sin^2(mk_z)
\end{eqnarray}

At the critical temperature T$_c$ $\vec{d}_\kappa$=0 and Eq.(\ref{DG2})
simplifies to

\begin{eqnarray}
\label{FUN}
F(T_c)&=&N(0)\int_0^{\omega_-^c}
\frac{\tanh\frac{1}{2}\beta_c\xi}{\xi}d\xi=
N(0)\ln(A\beta_c\omega_-^c)\nonumber\\
G_m[0]&=&\bar{\alpha}_mF(T_c)
\end{eqnarray}

Here A=1.13 and $\bar{\alpha}_m$ is the average of $\alpha_m(k_z)$ in the
interval [0,$\pi$]. One has $\bar{\alpha}_m$=1 for $\ka$=0, m=0 and
$\bar{\alpha}_m=\frac{1}{2}$ for $\ka$= 0,u,v,w and m$\geq$1. This leads to the
weak coupling BCS- formula for the transition temperature:

\begin{equation}
\label{TCE}
T^m_{c\kappa} = 1.13(\omega_-^c)\exp (-\frac{1}
{\bar{\alpha}_mN(0)|\bar{V}_\kappa^m|})
\end{equation}

\begin{figure}
\centerline{\psfig{figure=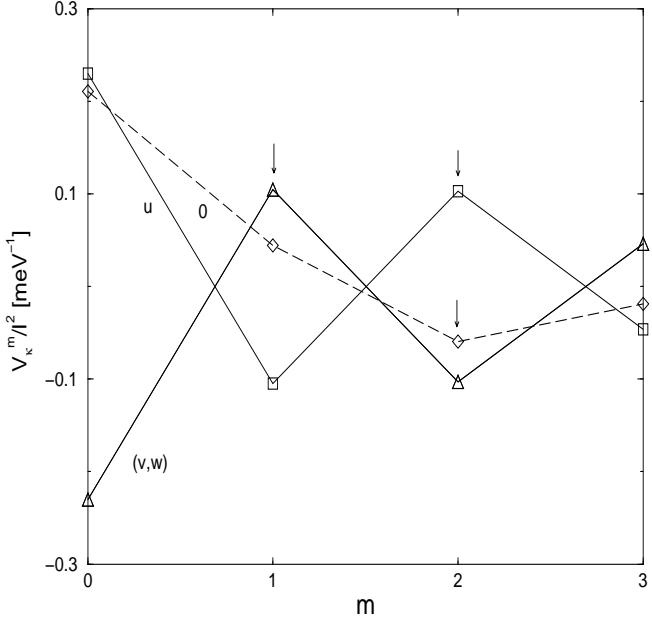,height=8.5cm,width=8.5cm,angle=-90}}
\vspace{1cm}
\caption{Fourier components $\bar{V}^m_{\kappa}$ of
$\bar{V}_{\kappa}$(q$_z$) according to Eq.(\ref{FOU}). dashed line (diamonds):
$\kappa$=0, full lines: $\kappa$=v,w (degenerate,triangles) and
$\kappa$=u(squares). The Fourier components with m$\leq$2 which lead
to a nonzero T$_{c\kappa}^m$ in Eq.(\ref{TCE}) have been designated by
arrows. The odd parity (v,w) component (m=1) is the most favorable.}
\label{FIGPOF}
\end{figure}

For T$_{c\ka}^m>$0 the conditions $\bar{V}_0^m<$0 for the even d$_0^m$
($\ka=0, m\geq 0$) singlet and  $\bar{V}_{\ka}^m>$0 ($\ka=u,v,w, m\geq 1$)
for odd parity d$_{\ka}^m$- states  must be satisfied. This difference in the
required sign of the Fourier components  $\bar{V}_0^m$ can again be
traced back to the decomposition in Eq.(\ref{FOU}). The Fourier
components in Eq.(\ref{FOU}) with m= 0-4 are shown in
Fig.\ \ref{FIGPOF}. Within the weak coupling approach the sc phase realised is
the one with the highest T$_{c\ka}^m$, i.e. the highest
$\bar{V}_{\ka}^m>$0 ($\ka$=u,v,w) or $-\bar{V}_0^m>$0 ($\ka$=0)
value. From the calculated values in Fig.\ \ref{FIGPOF} we notice that the
degenerate $\bar{V}_{\ka=v,w}^{m=1}$ is the most favorable which is
slightly larger than $\bar{V}_{\ka=u}^{m=2}$. The former corresponds
to an odd parity doublet

\begin{equation}
\label{ODD}
\vec{d}(k_z)=(d_v^1sink_z,d_w^1sink_z)
\end{equation}

with $|d_v^1|=|d_w^1|$. It has node lines of the gap at k$_z$=0. As
mentioned before this degeneracy is accidental in the present model
due to the neglect of AF ordering and therefore there is no symmetry
determined phase relation beween the d$_v^1$ and d$_w^1$
amplitudes. The second slightly less favorable odd parity state
d$_u^2\sin2k_z$ is nondegenerate and has line nodes at
k$_z$=0, $\frac{\pi}{2}$

An important conclusion from this analysis is that for the present
pure magnetic exciton model the singlet state 

\begin{equation}
\label{EVE}
d_0(k_z)=d_0^1\cos k_z
\end{equation}

is not stable because V$_0^1>$0 is repulsive instead of attractive as
required in this channel. This is due to the fact that the potential
$\bar{V}_0(q_z)$ in Fig.\ \ref{FIGPOZ}(dashed line) has its maximum
{\em not} at the PM zone boundary (q$_z=\pi$) but rather close to
q$_z=0.4\pi$. This behaviour can be traced back to the form of
V$_0(\vq)$ in Eq.(\ref{VIR}) which shows that the denominator
2$\Delta'-\omega_-(\vq)$ and not the exciton mode energy
$\omega_-(\vq)$ which leads to the maximum at q$_z=\pi$ for the 
odd parity state potentials $\bar{V}_{\ka}(q_z)$ at q$_z=\pi$ in the
PMBZ. This result is robust against the details of the exciton
exchange model.

\subsection{Effect of AF order on superconducting gap equations}

In the previous treatment the influence of AF order on the pair
potential was largely neglected. Sofar it only enters through the
renormalized singlet-singlet splitting $\Delta'$ in the magnetic
exciton dispersion which is of minor importance. A much
more pronounced effect is due to the competition of sc pair formation and
reconstruction of quasiparticle states at the FS by AF long
range order. This leads to an important modification of the effective
pair potential for wave vectors close to the AF ordering vector
$\vQ$. This is also known from AF phonon mediated superconductors such
as the borocarbides \cite{Morosow96,Amici00}. The application of this
idea to the present case is formally similar but more involved because
of the more general possible pair states as compared to conventional
electron-phonon singlet superconductors treated in the above
references. In this section we show how AF order modifies the magnetic
exciton pair potential of Eq.(\ref{FIN}) and the effect on the sc states as
discussed above.

\subsubsection{Transformation to magnetic Bloch states}

The AF order parameter

\begin{equation}
\label{SDW}
\la\vec{J}_i\ra =\la J\ra_0\hat{x}\cos(\vQ\vec{R}_i)
\end{equation}

leads to an additional periodic potential for the conduction electrons
due to H$_{cf}$ in Eq.(\ref{CF2}) which has twice the period (2c) of the
lattice potential (c). It is given by

\begin{equation}
\label{HSD}
H_{cf}^{mf}=I_0(g-1)\la J\ra_0\sum_{\vk}(c_{\vk+\vQ\da}^\dagger c_{\vk\ua}+
c_{\vk-\vQ\ua}^\dagger c_{\vk\da})
\end{equation}

and leads to a reconstruction of the conduction band states (see
e.g. Refs.[\onlinecite{Amici00,Amici99}]) which may be described by a
Bogoliubov transformation to the new {\em magnetic} Bloch states of
the AF lattice given by

\begin{eqnarray}
\label{BOG}
a_{\vk 1}^\dagger&=&
\al_{\vk}c_{\vk\ua}^\dagger+\be_{\vk}c_{\vk+\vQ\da}^\dagger\nonumber\\
a_{\vk+\vQ 2}^\dagger&=&
\be_{-\vk-\vQ}c_{\vk\ua}^\dagger+\al_{-\vk-\vQ}c_{\vk+\vQ\da}^\dagger
\end{eqnarray}

The $\al_{\vk}$, $\be_{\vk}$ are the real {\em magnetic} Bogoliubov
coefficients \cite{Morosow96,Amici00} below T$_N$ (not to be confused with
u$_{\vk}$,v$_{\vk}$ which transform to the superconducting
quasiparticles below T$_c$). The effect of AF on the pair potential
Eq.(\ref{FIN}) is included by expressing the latter with the transformed pair
operators $\vec{B}_{\vk}$=$\{a_{\vk\sigma}a_{-\vk\sigma'},
(\sigma\sigma')= 11,12,21,22\}$ instead of the pair operators 
 $\vec{b}_{\vk}$=$\{c_{\vk\sigma}c_{-\vk\sigma'}, (\sigma\sigma')=
\ua\ua,\ua\da,\da\ua,\da\da\}$ in the paramagnetic state. This
transformation $\vec{b}_{\vk}$=T$_{\vk}\vec{B}_{\vk}$ is furnished via
the symmetric 4$\times$4- matrix 

\begin{equation}
\label{TRA}
T_{\vk}=
\left(\matrix{
\al_{\vk}^2 &\al_{\vk}\be_{\vk} &\al_{\vk}\be_{\vk} &\be_{\vk}^2 & \cr
\al_{\vk}\be_{\vk} &\al_{\vk}^2 &\be_{\vk}^2 &\al_{\vk}\be_{\vk} & \cr
\al_{\vk}\be_{\vk} &\al_{\vk}^2 &\be_{\vk}^2 &\al_{\vk}\be_{\vk} & \cr
\al_{\vk}^2 &\al_{\vk}\be_{\vk} &\al_{\vk}\be_{\vk} &\be_{\vk}^2 & \cr
}\right)
\end{equation}
 
Then the effective pairing Hamiltonian transforms into ($\vq=\vk'-\vk$)

\begin{equation}
\label{HBG}
H_{eff}=\frac{1}{2}\sum_{\vk\vk'}\vec{b}_{\vk'}^\dagger
\tensor{V}_{eff}(\vq)\vec{b}_{\vk}=
\frac{1}{2}\sum_{\vk\vk'}\vec{B}_{\vk'}^\dagger
\tensor{\hat{V}}_{eff}(\vq)\vec{B}_{\vk}
\end{equation}

using the projector representation of Eq.(\ref{FIN}) the effective pair
potential may be written as 

\begin{eqnarray}
\label{VBO}
\tensor{\hat{V}}_{eff}(\vq)&=&
T_{\vk'}\tensor{V}_{eff}(\vq)T_{\vk}^\dagger \nonumber\\
&=&\sum_{\ka}V_{\ka}(\vq)T_{\vk'}P_{\ka}T_{\vk}^\dagger
=\sum_{\ka} V_{\ka}(\vq)\hat{P}_{\ka}
\end{eqnarray}

As long as $\vk,\vk'$
are far away from the AF Bragg planes $\vk=\pm\frac{1}{2}\vQ$ one has
T$_{\vk}$,T$_{\vk'}\simeq$ 1 and $\hat{P}_{\ka}\simeq P_{\ka}$
still projects to the eigenstates of Eq.(\ref{PSI}). However close to the
Bragg planes  $\hat{P}_{\ka}\neq P_{\ka}$ and
$\tensor{\hat{V}}_{eff}(\vq)$ and hence the gap equations will be
strongly modified. Explicitly we have

\begin{eqnarray}
\label{PRT}
\hat{P}_0&=&a_{\vk'}a_{\vk}P_0\nonumber\\
\hat{P}_u&=&a_{\vk'}a_{\vk}P_u\nonumber\\
\hat{P}_v&=&P_v+ b_{\vk'}b_{\vk}P_w 
+b_{\vk'}P_{vw}+b_{\vk}P_{wv}\\
\hat{P}_w&=&P_w+ b_{\vk'}b_{\vk}P_v 
+b_{\vk'}P_{vw}+b_{\vk}P_{wv}
\nonumber
\end{eqnarray}

Here we defined the transfer operators P$_{vw}$= P$_{wv}^{\dagger}$=
$|\psi_v\ra\la\psi_w|$ between different pair states. this means that AF
order mixes $\psi_v^{\vk}$, $\psi_w^{\vk}$ states for wave vectors
close to the Bragg planes, contrary to $\psi_0^{\vk}$, $\psi_u^{\vk}$
which stay in the same subspace. The coefficients in Eq.(\ref{PRT})
are related to the magnetic Bogoliubov coefficients via 

\begin{eqnarray}
\label{BRA}
a_{\vk}&=& \al_{\vk}^2-\be_{\vk}^2
=[1+\frac{(I\la S_x\ra)^2}{(\epsilon_{\vk}-\epsilon_{\vk+\vQ})^2}]
^{-\frac{1}{2}}\nonumber\\
b_{\vk}=(1-a_{\vk}^2)^\frac{1}{2}
&=&2\al_{\vk}\be_{\vk}=\frac{I\la S_x\ra}
{|\epsilon_{\vk}-\epsilon_{\vk+\vQ}|}(\al_{\vk}^2-\be_{\vk}^2)
\end{eqnarray}

Where I$\la S_x\ra$=I$_0$(g-1)$\la J\ra_0$

\subsection{Modified gap equations in the AF state}

The above relations enable us to derive the modified sc gap equations
in the presence of AF order. The effective pairing of
Eqs.(\ref{HBO},\ref{VBO}) leads to gap equations analogous to
Eq.(\ref{GA1}) for electrons in magnetic Bloch states:

\begin{equation}
\label{MGA}
\vec{\hat{\Delta}}_{\vk}= 
-\sum_{\vk'}\hat{V}_{eff}(\vq)\vec{\hat{\Delta}}_{\vk'}F_{\vk'}
\end{equation}

Multiplikation with a projektor P$_\ka$ from the left and using the
identities  P$_v$P$_{vw}$=P$_{vw}$, P$_v$P$_{wv}$ =0 etc. we
arrive at the modified gap equations for $\ka$=0,u:

\begin{eqnarray}
\label{MDG}
\hat{d}_{\ka}^{\vk}&=&
-\sum_{\vk'}[a_{\vk'}^2V_{\ka}(\vq)]\hat{d}_{\ka}^{\vk'}F_{\vk'}\nonumber\\
E_{\vk}&=&[\xi_{\vk}^2+d_{\ka}^{\vk 2}]^{\frac{1}{2}}\\
d_{\ka}^{\vk}&=&a_{\vk}\hat{d}_{\ka}^{\vk}
=(\al_{\vk}^2-\be_{\vk}^2)\hat{d}_{\ka}^{\vk}\nonumber
\end{eqnarray}

This shows that for the  $\psi_0^{\vk}$, $\psi_u^{\vk}$ pair states
the effective pair potential in the presence of AF is given by

\begin{equation}
\label{VH2}
\hat{V}_{\ka}(\vk-\vk')=
(\alpha_{\vk}^2-\beta_{\vk}^2)V_{\ka}(\vk-\vk')
(\alpha_{\vk'}^2-\beta_{\vk'}^2)
\end{equation}

which vanishes at the Bragg planes
$\vk=\pm\frac{1}{2}\vQ$, i.e. states which
are connected by the AF ordering vector do not contribute to sc
pairing. This leads to the well known depression but usually not
destruction of superconductivity  by AF order, see
e.g. Ref.(\onlinecite{Amici00}). Far from the
magnetic Bragg planes $\hat{V}_{\ka}(\vk-\vk')\simeq
V_{\ka}(\vk-\vk')$. Furthermore the gap function entering the
quasiparticle energy E$_{\vk}$ is d$_{\ka}^{\vk}$ which has an
additional zero for $\vk=\pm\frac{1}{2}\vQ$ due to the prefactor
$(\al_{\vk}^2-\be_{\vk}^2)$. This factor does not change sign at
$\vk=\pm\frac{1}{2}\vQ$ and therefore does not lead to an additional
node line. The modified gap equations can now be derived in a similar way
for $\ka$=u,v:

\begin{eqnarray}
\label{DMA}
\left(\matrix{
d_v^{\vk} \cr
d_w^{\vk} \cr
}\right)&=&-\sum_{\vk'}
\left(\matrix{
V_v(\vq)          & ib_{\vk}V_v(\vq) & \cr
-ib_{\vk}V_w(\vq) & V_w(\vq)         & \cr
}\right)
\left(\matrix{
d_v^{\vk'} \cr
d_w^{\vk'} \cr
}\right)F_{\vk'}\nonumber\\
&&
\end{eqnarray}

Obviously the two states (d$_v^{\vk}$,d$_w^{\vk}$) are mixed at the presence
of AF order. According to Eq.(\ref{VIR}) V$_v$= V$_w\equiv$ V$_D$ and
Eq.(\ref{DMA}) can easily be diagonalised. One finally obtains the gap
equations for the new eigenstates ($\kappa$=+,-):

\begin{eqnarray}
\label{DMO}
\hat{d}_\pm^{\vk}&=&-\sum_{\vk'}[(1\mp b_{\vk'})V_D(\vq)]
\hat{d}_\pm^{\vk'}F_{\vk'}\nonumber\\
E_{\vk}&=&[\xi_{\vk}^2+d_{\ka}^{\vk 2}]^{\frac{1}{2}}\\
d_\pm^{\vk}&=&(1\mp b_{\vk})^\frac{1}{2}\hat{d}_\pm^{\vk}
=(1\mp 2\alpha_{\vk}\beta_{\vk})^\frac{1}{2}\hat{d}_\pm^{\vk}\nonumber\\
d_\pm^{\vk}&=&\frac{1}{\sqrt{2}}(d_v^{\vk}\pm id_w^{\vk})
\nonumber
\end{eqnarray}

These equations for $\psi_+, \psi_-$ pair states are formally
identical to those for $\psi_0, \psi_u$ in Eq.(\ref{MDG}) however with
a different effective pair potential

\begin{equation}
\label{VH1}
\hat{V}_\pm(\vk-\vk')= 
(1\mp b_{\vk})^\frac{1}{2}V_D(\vk-\vk')(1\mp b_{\vk'})^\frac{1}{2}
\end{equation}

At the Bragg planes b$_{\vk}\simeq 1$, this implies that the pair
potential V$_+$ for d$_+^{\vk}$ is reduced to zero, similar as in the
case $\ka$=0,u but for d$_-^{\vk}$ the pair potential V$_-$ is
enhanced in this region of k- space, therefore AF order supports the sc
pairing in the d$_-^{\vk}$ state.

As for the PM case in Sect. (V.A) we will now consider the
simple situation of the FS sheet with cylindrical symmetry and small
dispersion along c shown in the inset of Fig.\ \ref{FIGOPA}. Then
again the gap equations can be decomposed into Fourier components
which satisfy Eq.(\ref{DG2}) as in the PM case. However
the form factor $\alpha_m(k_z)$ and the quasiparticle energy
E($\xi$,k$_z$) are now different because they include the effect of AF
order. One obtains from Eqs.(\ref{MDG},\ref{DMO})

\begin{eqnarray}
\label{AFG}
\ka=0:\;\;\;
\alpha_m(k_z)&=&g(k_z)\cos^2(mk_z)\nonumber\\
E(\xi,k_z)&=&
[\xi^2+g(k_z)(\sum_nd_{\ka}^n\cos nk_z)^2]^{\frac{1}{2}}\nonumber\\ 
\ka=u:\;\;\;
\alpha_m(k_z)&=&g(k_z)\cos^2(mk_z)\nonumber\\
E(\xi,k_z)&=&
[\xi^2+g(k_z)(\sum_nd_{\ka}^n\sin nk_z)^2]^{\frac{1}{2}}\\ 
\ka=\pm :\;\;\;
\alpha_m(k_z)&=&h_{\ka}(k_z)\cos^2(mk_z)\nonumber\\
E(\xi,k_z)&=&
[\xi^2+h_{\ka}(k_z)(\sum_nd_{\ka}^n\sin nk_z)^2]^{\frac{1}{2}}
\nonumber
\end{eqnarray}

Here we defined g(k$_z$)=a$_{\vk}^2$=$(\al_{k_z}^2-\be_{k_z}^2)^2$ and
h$_\pm$(k$_z$)= (1$\mp 2\al_{k_z}\be_{k_z}$). Using
Eq.(\ref{BRA}) one can derive

\begin{eqnarray}
\label{AFP}
g(k_z)&=&\frac{\cos^2k_z}{\lambda_{AF}^2+\cos^2k_z}\nonumber\\
h_\pm(k_z)&=&=1\mp[1-g(k_z)]^\frac{1}{2}
\end{eqnarray}

The functions g(k$_z$), h$_\pm$(k$_z$) with the parameter
$\lambda_{AF}\ll$ 1 given below accounts for the influence of AF order
on the sc gap via the FS reconstruction caused by H$_{cf}^{mf}$ in
Eq.(\ref{HSD}). Specifically a finite $\lambda_{AF}$ will spit the
degenerate doublet pair state (d$_v^{\vk}$,d$_w^{\vk}$) of the
PM case.

\subsection{Numerical solution of gap equations}

We now investigate some of the possible solutions of the gap equations
quantitatively. It must be kept in mind that this work is based on a
model theory which cannot make absolute predictions for the
experimentally realised gap function. Its main purpose
is a better understanding of the physics of a novel type of pairing
mechanism leading to the magnetic exciton mediated superconductivity. 

\begin{figure}
\centerline{\psfig{figure=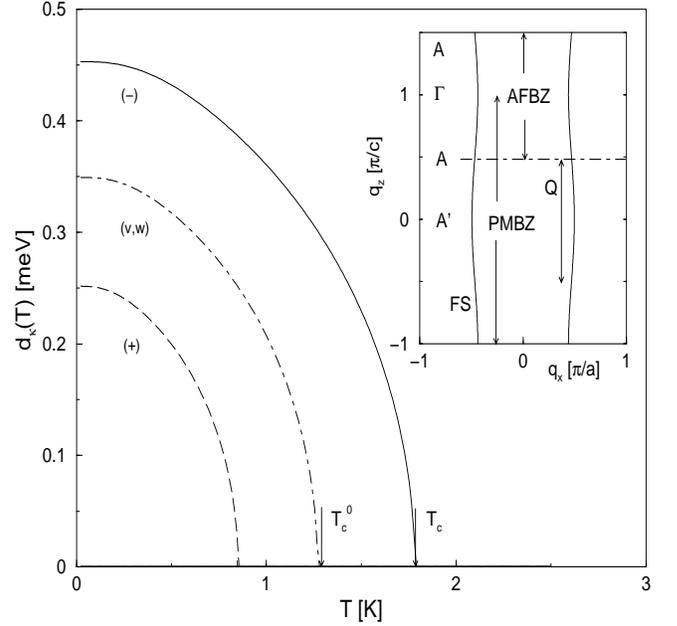,height=8.5cm,width=8.5cm,angle=-90}}
\vspace{1cm}
\caption{Temperature dependence of gap function amplitudes
d$_-$(T) (full line) and d$_+$(T) (dashed line) with AF order
effect included ($\lambda_{AF}$ according to
Eq.(\ref{LAP})). $\Delta\lambda_{sc}$ =4.3 meV is chosen such that
T$_c$=1.8 K. dashed-dotted line: AF moment effect suppressed
($\lambda_{AF}\equiv$0), with the same $\lambda_{sc}$, the transition
temperature is then T$^0_c$=1.27 K.
{\em inset}: Schematic view of PM Fermi surface (FS) described by
Eq.(\ref{DIS})
with PM and AF Brillouin zones indicated. $\Gamma$ and A' are the AFBZ
and PMBZ zone centers respectively. States on the FS connected by the
AF ordering wave vector $\vQ$ are reconstructed by the AF order.}
\label{FIGOPA}
\end{figure}

We first give estimates of the physical parameters involved in the
present superconducting model for \U in addition to those already
discussed in Sect.(III) for the magnetic exciton model itself. According
to Ref.[\onlinecite{Knoepfle96}] the Fermi level is in the middle of a
peak which we approximate by a square DOS of width W=2E$_F$=0.88
eV. Assuming one conduction electron per U site we have
N(0)=$\epsilon_F^{-1}$ = 0.0023 meV$^{-1}$ for the model DOS. The small
bandwidth W$_\parallel\ll$W of the k$_z$ dispersion is given by
W$_\parallel$=4t$_\parallel\simeq$ 2E$_F$(A$_c^{max}$/A$_c^{min}$-1)
where the last approximate relation is obtained assuming a free
electron like dispersion $\perp$ c and relating the change of the FS
cylinder cross section A$_c$ in the PMBZ along c to the hopping
t$_\parallel$. The deviation of the ratio A$_c^{max}$/A$_c^{min}$ from
1 characterizes the amount of 'corrugation' of the FS cylinder along c as
shown in Fig.\ \ref{FIGOPA}, its value is obtained from dHvA
experiments\cite{Inada99} as A$_c^{max}$/A$_c^{min}\simeq$  1.24 in
the AFBZ. This leads to an estimate W$_\parallel$= 0.21eV or
(W$_\parallel$/W)= 0.24 consistent with the assumption made above. 
The interaction H$_{cf}$ between localized 5f and conduction electrons
with strength I=$\alpha$I$_0$(g-1) has two competing effects which are
characterized by the dimensionless coupling constants

\begin{equation}
\label{LAP}
\lambda_{sc}=\frac{N(0)I^2}{\Delta}\;\;\;
\lambda_{AF}=\frac{(I/\alpha)\la J\ra _0}{4t_\parallel}
\end{equation}

Whereas $\lambda_{sc}$ characterizes the strength of the pair potential due to
magnetic exciton exchange, $\lambda_{AF}$ is associated with the
effect of AF on the sc pair states. Since both are determined by the
interaction constant I they are not independent and one has the relation 
 
\begin{equation}
\label{LAF}
\lambda_{AF}=\frac{1}{2}\frac{\la J\ra_0}{\alpha}\frac{W}{W_\parallel}
[N(0)\Delta]^\frac{1}{2}\lambda_{sc}^\frac{1}{2}
\end{equation}

The quantities W$_\parallel$/W= 0.24 and
[N(0)$\Delta$]$^\frac{1}{2}$= 0.12 are estimated within our model  so
that only one independent parameter $\lambda_{sc}$ remains which will
be fixed to achieve the proper T$_c$= 1.8 K. The solution of the gap
equations Eq.(\ref{DG2})  for the AF case of Eq.(\ref{AFG}) proceeds
by iteration. As discussed in Sect.(V.A) the most favorable pair
state, judging from the irreducible potentials in Fig.\ \ref{FIGPOF}
is the odd parity doublet given by Eq.(\ref{ODD}). 

The temperature variation of its sc amplitudes d$_+$(T), d$_-$(T)
(i.f. the index m=1 is suppressed) is shown in Fig.\ \ref{FIGOPA} for
two cases:
(a) neglect of AF order, setting $\lambda_{AF}\equiv$0 arbitrarily
then both representations are accidentally degenerate (b) including the
effect of AF with the proper $\lambda_{AF}$ according to Eq.(\ref{LAF}): the
degeneracy is lifted due to the finite $\lambda_{AF}$ and the
nondegenerate d$_-^{\vk}$ state describes the true sc phase in the AF
background. It is prefered as compared to d$_+^{\vk}$ due to its gap
enhancement factor (1+b$_{\vk}$) in Eq.(\ref{DMO}). 
The iteration procedure used for the
calculation of d$_{\ka}$(k$_z$) in Eq.(\ref{DG2}) allows for the
appearance of different Fourier components below T$_{c\ka}$ which
correspond to V$^n_{\ka}$ irreducible potentials which are smaller
than the maximum
one with n=m. Their appearance would lead to a change of the profile of 
d$_{\ka}$(k$_z$) as function of temperature. For all cases studied
with the potentials of Fig.\ \ref{FIGPOF} such a case never appears,
therefore in the quasiparticle energies in Eq.(\ref{AFG}) the
summation over n can simply be replaced by the dominant term for n=m.

\begin{figure}
\centerline{\psfig{figure=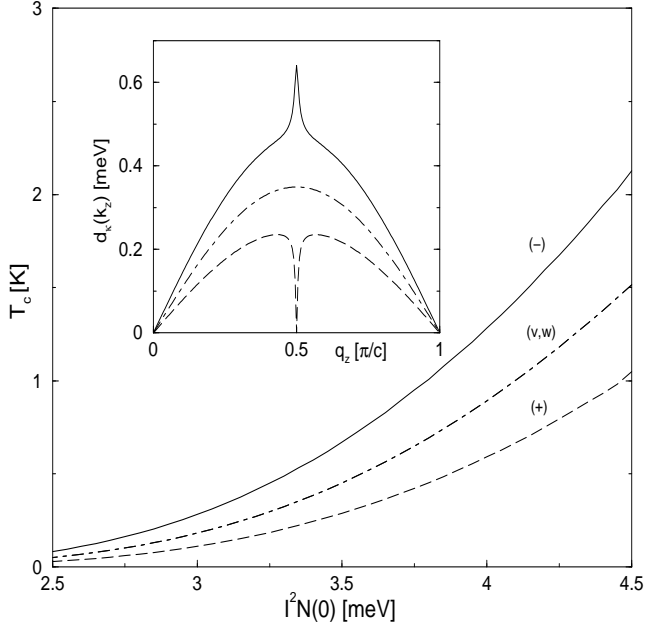,height=8.5cm,width=8.5cm,angle=-90}}
\vspace{1cm}
\caption{Variation of transition temperature with coupling strength
$\Delta\lambda_{sc}$=I$^2$N(0) for the (v,w) odd parity doublet states. With
AF order effect suppressed (($\lambda_{AF}\equiv$0) T$_c$'s are
degenerate (dash- dotted line). With effect of $\lambda_{AF}$
(Eq.(\ref{LAP})) included a splitting appears (full and dashed
lines). Therefore in the
presence of AF only the nondegenerate d$_-$(k$_z$) state of Eq.(\ref{DVK})
will be stable at any temperature below T$_c^v$ =1.8 K for
$\Delta\lambda_{sc}$=4.3 meV. {\em inset:} wave number (k$_z$-)
dependence of the gap functions d$_\pm$(k$_z$) and  d$_{v,w}$(k$_z$)
($\lambda_{AF}\equiv$0) at T= 0.05 K. The anomalies at the Bragg plane
$\frac{1}{2}Q_z$= $\frac{\pi}{2c}$ are due to the effect of AF order.}
\label{FIGTCD}
\end{figure}

In Fig.\ \ref{FIGTCD} it is shown how the suppression of T$_c$ and the
splitting of  d$_+^{\vk}$, d$_-^{\vk}$ representations evolves as function of 
 $\lambda_{sc}$ due to the effect of $\lambda_{AF}$ given in
Eq.(\ref{AFP}). AF order renders the d$_-^{\vk}$ state
more favorable. Once d$_-^{\vk}>$0 at T$_{c}$ the split- off
doublet component d$_+^{\vk}$ with its lower transition temperature will be
completely suppressed by the already finite condensation energy of the
d$_-^{\vk}$ state. Therefore in the AF background only a single
nondegenerate state 

\begin{equation}
\label{DVK}
d_-^{\vk}=d_-(T)(1+2\al_{\vk}\be_{\vk})^\frac{1}{2}\sin k_z
\end{equation}

will appear. The k$_z$- dependence of d$_\pm^{\vk}$ and d$_{(v,w)}^{\vk}$
($\lambda_{AF}\equiv$ 0) is illustrated in Fig.\ \ref{FIGTCD} (inset).
Here $\Delta\lambda_{sc}$= 4.3 meV ($\lambda_{sc}$=0.72), which after
Eq.(\ref{LAF}) corresponds to $\lambda_{AF}$=0.018, is fixed to
reproduce the physical value for T$_c$=1.8 K for d$_-^{\vk}$.
 
\section{A hybrid model including spin fluctuations}

It has been the aim of the previous analysis to investigate a new type
of pairing mechanism which is a consequence of the dual nature of 5f-
electrons in \U. The model employed in Eq.(\ref{HAM}) as an extreme case
without any repulsion (I$_c$=0) between 5f- conduction electrons of the
type given in Eq.(\ref{HIC}). Therefore only the new pairing mechanism via the
exchange of magnetic excitons is possible which was then studied for
its own sake. In a real 5f- compound one has to take into
account also the quasiparticle Coulomb repulsion (I$_c>$0) described
by Eq.(\ref{HIC}) which leads to a conventional spin fluctuation type part for
the total pairing potential. In this last section we therefore study a
hybrid model where both contributions are simply added, neglecting any cross
influence. We investigate the evolution of the pair potential and
favorable sc states as a function of the ratio $\rho$=(I$_c$/I) which
controls the relative strength of these contributions. The pure
magnetic exciton model studied before corresponds to $\rho$=0. The two
contributions of Eqs.(\ref{HIC},\ref{CF2}) to the total pair potential are
determined by the RPA static conduction electron susceptibility
$\chi(\vq)$ and the exciton mode energy $\omega_-(\vq)$
respectively. Due to Eqs.(\ref{JTO},\ref{VXC}) they are related via

\begin{equation}
\label{HY1}
I^2\chi_0(\vq)=-[\omega_-(\vq)+\alpha^2\la S\ra J'_{ff}(\vq)]+\omega_0
\end{equation}

with $\omega_0=\Delta'+I^2\bar{\chi}$,
$\bar{\chi}=\frac{1}{N}\sum_{\vq}\chi(\vq)$. Since the superexchange
J'$_{ff}(\vq)$ is unknown we restrict to the simplest possible case
assuming J'$_{ff}(\vq)\simeq$0, then $\chi_0(\vq)$ is directly
determined by the magnetic exciton dispersion $\omega_-(\vq)$ up to a
constant $\omega_0$ which has to satisfy the constraint

\begin{equation}
\label{HY2}
I^2\chi_0(\vq)=\omega_0-\omega_-(\vq)\geq0
\end{equation}

Then, using Eqs.(\ref{V01},\ref{VIR}) the total pair potential of the
hybrid model is given by 

\begin{eqnarray}
\label{HYP}
V_s&=& I^2[\frac{3}{2}\rho^2\chi(\vq)+\frac{1}{\omega_R(\vq)}]\nonumber\\
V_u&=&-I^2[\frac{1}{2}\rho^2\chi(\vq)-\frac{1}{\omega_-(\vq)}]\\
V_v=V_w&=&-I^2[\frac{1}{2}\rho^2\chi(\vq)+\frac{1}{\omega_-(\vq)}]
\nonumber
\end{eqnarray}

where the RPA conduction electron susceptibility of Eq.(\ref{SUS}) is
obtained as

\begin{equation}
\label{HMO}
\rho^2\chi(\vq)=\frac{(\frac{\rho}{I})^2[\omega_0-\omega_-(\vq)]}
{1-\frac{\rho}{I}[\omega_0-\omega_-(\vq)]}
\end{equation}

\begin{figure}
\centerline{\psfig{figure=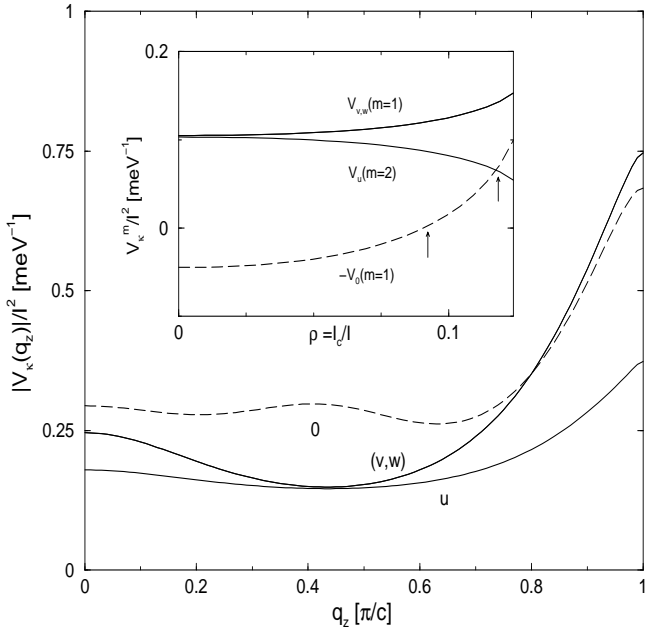,height=8.5cm,width=8.5cm,angle=-90}}
\vspace{1cm}
\caption{Effective pair potentials (absolute values) for the hybrid
model for even (0) and odd (u,v,w) parity for
$\rho$= I$_c$/I =0.124. {\em inset:} most important Fourier components as
function of model parameter $\rho$. With increasing  $\rho$
(increasing spin fluctuation contribution) the even parity singlet
($\kappa$=0, m=1) becomes stable at $\rho$ = 0.097 (first arrow) and
then more favorable than the odd parity u state (second arrow) but stays less
favorable than the (v,w) odd parity doublet up to the maximum $\rho$= 0.124.}
\label{FIGHYB}
\end{figure}

For the hybrid model of Eq.(\ref{HYP}) one can now calculate the q$_x$, q$_y$-
averaged potentials V$_\kappa$(q$_z$) as in Eq.(\ref{VAV}) where the
bar will now be suppressed. They are shown in
Fig.\ \ref{FIGHYB} as function of q$_z$ for parameters $\omega_0$,
$\rho$ chosen such that Eq.(\ref{HY2}) is fulfilled and $\chi(\vq)$
stays paramagnetic, i.e. nonsingular. One observes that the singlet
potential (V$_s$) has now also a pronounced maximum at $\vQ$
comparable to the potentials V$_{v,w}$ for the odd parity doublet
state. Note that for the hybrid
model V$_u\neq$ -V$_{v,w}$ and therefore three potential functions are
present in Fig.\ \ref{FIGHYB} contrary to the situtation in the pure magnetic
exciton model of Fig.\ \ref{FIGPOZ}. The stability of the singlet sc
state should now be increased as $\rho$ increases. This can clearly be
seen by plotting the relevant Fourier components V$_\kappa^m$ as
function of
increasing $\rho$, i.e. increasing the spin fluctuation contribution
to the pairing. This is done for the unstable singlet state in Fig.\
\ref{FIGPOF}
(dashed curve, m=1) versus the two most favorable odd parity states
(upper arrows in Fig.\ \ref{FIGPOF}) and the result is shown in Fig.\
\ref{FIGHYB}. For
$\rho\geq$ 0.09 one has -V$_s>$0 and the singlet state becomes
possible. For even larger $\rho\geq$ 0.12 it becomes more stable than 
the d$_u(\vk)$ (m=2) odd parity state and then -V$_s$ rapidly approaches
the potentials V$_{v,w}$ for the doublet state for the largest value
of $\rho$. At this value the model $\chi(\vq)$ in Eq.(\ref{HY2}) becomes
singular indicating an AF instability of the conduction electron
system. For the hybrid model with J'$_{ff}(\vq)$=0 and $\chi(\vq)$ given
by Eq.(\ref{HMO}) the odd parity state is therefore always the stable one,
however the singlet state d$_0$(k$_z$) becomes increasingly favorable
as the spin fluctuation part increases. In principle it is possible that
in the general case with the (unknown) J'$_{ff}(\vq)$ included in
Eq.(\ref{HY1}) the hybrid model will prefer the singlet sc state
d$_0(\vk)$. Of course, if we neglect the magnetic exciton part
altogether and consider only the isotropic pure spin fluctuation model
then according to Eq.(\ref{V01}) V$_s\equiv$V$_0$=-3V$_{u,v,w}$, and provided
$\chi(\vq)$ has the maximum at $\vq=\vQ$, the singlet even parity
sc state  d$_0$(k$_z$)= d$^1_0\cos k_z$ is the stable one. 

\section{Summary and Outlook}
In this work a dual model of localised and itinerant 5f- electrons was
studied as a generic model for superconductivity in uranium- compounds and
applied to \U. The 
dual nature of 5f electrons in this compound was concluded from
experimental evidence, notably susceptibility and normal state Knight shift
measurements. The localized electrons were assumed to be in a 5f$^2$
configuration split by the CEF into singlet ground and excited states
separated by an energy $\Delta$=6 meV. The conduction electrons form
band states whose main FS sheet can be described by a corrugated
cylinder aligned along the c- axis. Due to RKKY and superexchange
interactions the CEF excitation at $\Delta$ is broadened into a dispersive band
of magnetic excitons which become soft at the Neel temperature T$_N$
where an induced AF moment appears in the singlet ground state. 
These magnetic excitons where found in inelastic neutron scattering
experiments \cite{Mason97,Bernhoeft98,Sato97} and a theoretical
explanation of their dispersion has been given within the AF singlet
ground state model. Furthermore a comparison with tunneling
experiments \cite{Jourdan99} has suggested that the exchange of
magnetic excit ns between conduction electrons leads to a new type of
pairing mechanism which is
responsible for superconductivity in \U. The consequences of this
hypothesis have been analyzed in detail within the dual model. The new
magnetic exciton mediated pair potential
was derived within a diagrammatic RPA type approach and compared to
the results of the well known spin fluctuation theory pair potential
caused by anti- paramagnon exchange. It was found that the former
manifestly violates rotational symmetry in spin space contrary to the
latter which is composed of spin singlet and triplet contributions due
to rotational invariance. The new effective pair potential due to magnetic
exciton exchange has a lower symmetry with contributions from the even
singlet, another odd parity nondegenerate and an odd parity doublet
contribution. The projector representation defining the irreducible
pair potentials has been derived and
for the cylindrical FS sheet the effectively 1D gap equations have
been solved. The favoured representation is determined by the largest Fourier
coefficient of the in-plane averaged pair potential. In case of
AF magnetism is excluded it was found that the new mechanism favors
the odd parity doublet state which has equatorial node lines in the
gap function. The singlet state is generally not favored in the
present weak coupling version of the magnetic exciton mechanism. 
The influence of singlet ground state AF on sc leads to important
consequences: The doublet state is strongly split and the stable sc
state is a nondegenerate odd parity state given by Eq.(\ref{DVK}). 

A hybrid model including spin fluctuations has also
been considered. In the simplified case that superexchange between
localized 5f electrons can be neglected the spin fluctation pair
potential can also be obtained from the magnetic exciton
dispersion. As expected its admixture leads to a stabilization of the
singlet pair state with node lines at
$\pm\frac{1}{2}\vec{Q}$. However within the simplified model the above
mentioned odd parity state remains the sc ground
state. In a more general hybrid model which does not have a direct
connection between the two contributions to the pair potential an
even singlet sc ground state is possible which is favored
for the pure spin fluctuation model of the pair potential.
In this work the thermodynamic signatures of the proposed and possible
gap functions have not been discussed. Their detailed investigation is
an important aspect for future work.

Thermal conductivity measurements \cite{Chiao97,Hiroi97} point
to the existence of node lines in the gap in a plane parallel to the
hexagonal ab plane but sofar do not allow to give their position
ck$_z$ along the c axis. Recently it has been proposed
\cite{Maki00,Thalmeier01b} that the field-angle dependence of thermal
conductivity gives a unique possibility to determine the position of
the node lines. It has been shown \cite{Thalmeier01b} that the angle
dependence should be distinctly different for an odd parity gap function $\sin
k_z$ with k$_z$=0 node line as in Eq.(\ref{ODD}) and the even parity gap
function $\cos k_z$ with node line at  k$_z$=$\frac{\pi}{2}$ as in
Eq.(\ref{EVE}). It is generally
believed that the observation of a Knight shift below
T$_c$\cite{Feyerherm94,Tou95,Matsuda97} points to an singlet even parity gap
function as in Eq.(\ref{EVE}) which is not favored by the present form
of the magnetic
exciton model. However the Knight shift has also large contribution
from the AF ordered localised moments and the quantitative extraction
of the itinerant part is not unique. Therefore one has to await the
field- angle dependent thermal conductivity measurements to solve the
problem of the true node line position and parity of the order parameter.

The theory developed in this paper illustrates the basic mechanism of
Cooper pair formation via the magnetic exciton exchange in U-compounds
with dual (localised and itinerant) 5f electrons. In its context
differences to the conventional spin fluctuation mechanism and
general properties of the nonretarded pair potentials were investigated.
However one cannot expect fully quantitative predictions for \U from
such a simplified weak coupling model. Ultimately a strong
coupling theory with a retarded magnetic exciton mediated pair
potential is necessary for a realistic application of this new pairing
mechanism to \U and a theoretical approach to the question of stability
of even or odd parity gap function.

\section*{Acknowledgement}
The author would like to thank K. Maki, N. Sato, R. Shiina, A. N. Yaresko and
G. Varelogiannis for useful discussions.

\newpage

\appendix

\section{}
In this appendix we give the basic geometry of the conventional unit
cells in the direct (Wigner Seitz cell: WS) and reciprocal (Brillouin
zone: BZ) hexagonal lattice for the discussion of the magnetic exciton
dispersion. The BZ is rotated by
30$^\circ$ with respect to the WS cell. The relevant primitive lattice
vectors of the direct and reciprocal lattices are given, respectively, by

\begin{eqnarray}
\label{DIR}
\vec{a}&=&a(\frac{1}{2},-\frac{\sqrt{3}}{2},0)\nonumber\\
\vec{b}&=&a(\frac{1}{2}, \frac{\sqrt{3}}{2},0)\\
\vec{b}&=&c(0, 0, 1)\nonumber
\end{eqnarray}

\begin{eqnarray}
\label{REC}
\vec{a}^*&=&\frac{2\pi}{a}(1,-\frac{1}{\sqrt{3}},0)\nonumber\\
\vec{b}^*&=&\frac{2\pi}{a}(1,\frac{1}{\sqrt{3}},0)\\
\vec{b}^*&=&\frac{2\pi}{c}(0, 0, 1)\nonumber
\end{eqnarray}

Any wave vector can be written in this basis as $\vec{q}$=h$\vec{a}^*$+
k$\vec{b}^*$+ l$\vec{c}^*$. In reduced units of
$\frac{1}{a}$ and $\frac{1}{c}$, $\vec{q}$ is then given by

\begin{eqnarray}
\label{BZS}
\mbox{PMBZ}:~\vec{q}&=&2\pi(h+k,\frac{1}{\sqrt{3}}(h-k),l)\nonumber\\
\mbox{AFBZ}:~\vec{q}'&=&2\pi(h+k,\frac{1}{\sqrt{3}}(h-k),\frac{1}{2}l')
\end{eqnarray}

For the q$_z$ components the relation $l'=2l-1$ or
$l=\frac{1}{2}(l'+1)$ holds, i.e. the AFBZ is obtained from the PMBZ
through the shift by an AF wave vector $\vQ=(0,0,\pi)$ and
scaling by a factor $\frac{1}{2}$. The corresponding wave vectors
$\vq~'=\vq-\vQ$ of the symmetry points of the hexagonal AFBZ are then
given by

\begin{eqnarray}
\Gamma &: & \pi(0, 0, 0)\nonumber\\
M&:      & \pi(1,-\frac{1}{\sqrt{3}},0)\nonumber\\
K&:      & \pi(\frac{4}{3},0,0)\nonumber\\
A&:      & \pi(0,0,\frac{1}{2})\nonumber\\
A'&:     & \pi(0,0,1)\nonumber\\
L&:      & \pi(1,-\frac{1}{\sqrt{3}},\frac{1}{2})\nonumber\\
H&:      & \pi(\frac{4}{3},0,\frac{1}{2})\nonumber
\end{eqnarray}

In the AFBZ the symmetry point A' is equivalent to $\Gamma$ since it
is shifted by a reciprocal lattice vector $\vQ$.
\newpage

\section{}
The single- ion susceptibility tensor $\tensor{u}(\omega)$ which is a
necessary ingredient for the calculation of the magnetic exciton
dispersion from Eq.(\ref{RPA}) is obtained from the general linear
response expression for two operators A, B in a system of two levels
$\epsilon_\pm$ corresponding to CEF singlet eigenstates $|\pm\ra$ in
the molecular field:

\begin{eqnarray}
\label{LOC}
u_{BA}(\omega)&=&P(T)[\frac{M_{BA}^{+-}}{\Delta'+\omega}
+\frac{M_{BA}^{-+}}{\Delta'-\omega}] \nonumber\\
P(T)&=&2\la S\ra =\tanh \frac{\beta}{2}\Delta' \\
M_{BA}^{\alpha\beta}&=&\la\alpha|\hat{B}|\beta\ra\la\beta|\hat{A}|\alpha\ra
\nonumber
\end{eqnarray}

Here $\la S\ra\equiv\la S_z\ra$, $\Delta'=\epsilon_+-\epsilon_-$ is the
CEF excitation energy of
Eq.(\ref{DEL}), $\alpha, \beta= \pm$ refer to the mf- eigenstates in the
ordered phase (Eq.(\ref{MFS})) and $\hat{A}$=A-$\la A\ra$, $\hat{B}$=A-$\la
B\ra$. For A,B= J$_\alpha$ ($\alpha$ =x,y,z) the form of the 3$\times$3
dipolar cartesian susceptibility tensor $\tensor{u}(\omega)$ is too
cumbersome for solving the secular equation for the exciton modes. This
may be remedied by performing a real space rotation
(x,y,z)$\rightarrow$(x',y,z') around the y- axis by an angle
$\theta_r$ such that it exactly compensates the effect of the rotation
in state space which transforms to the eigenstates $|\pm\ra$ of the
mf- Hamiltonian Eq.(\ref{MFS}). In this basis one has

\[\tilde{J}_x=\frac{1}{2}\alpha 
\left(\matrix{
-b & a\cr
 a & b\cr}\right)\]

\begin{equation}
\label{JTI}
\tilde{J}_y=J_y=\frac{1}{2}\alpha 
\left(\matrix{
0 &-i\cr
i & 0\cr}\right)
\end{equation}

\[\tilde{J}_z=\frac{1}{2}\alpha 
\left(\matrix{
a & b\cr
b &-a \cr}\right)\]

with a=$\cos 2\theta$=[1+$\gamma'^2]^{-\frac{1}{2}}$, b=$\sin 2\theta=
\gamma 'a$. Performing the counter- rotation leads to
(u$_r=\cos\theta_r$, v$_r=\sin\theta_r$)

\begin{equation}
\label{JR1}
J'_x=u_r\tilde{J}_x-v_r\tilde{J}_z,~ 
J'_y=\tilde{J}_y,~ 
J'_x=u_r\tilde{J}_x-v_r\tilde{J}_z
\end{equation}

defining $\phi=2\theta$ this can be written

\[J'_x=\frac{1}{2}\alpha 
\left(\matrix{
-\sin(\theta_r+\phi)&\cos(\theta_r+\phi) \cr
 \cos(\theta_r+\phi)&\sin(\theta_r+\phi) \cr}\right)\]

\begin{equation}
\label{JR2}
J'_y=J_y=\frac{1}{2}\alpha 
\left(\matrix{
0 &-i\cr
i & 0\cr}\right)
\end{equation}

\[J'_z=\frac{1}{2}\alpha 
\left(\matrix{
\cos(\theta_r+\phi) & \sin(\theta_r+\phi)\cr
\sin(\theta_r+\phi) &-\cos(\theta_r+\phi)\cr}\right)\]

By choosing $\theta_r\equiv -\phi= -2\theta$ the two transformations
compensate and J'$_x$=J$_x$, J'$_y$=J$_y$, J'$_z$=J$_z$. Therefore in
the rotated (x',y',z')- coordinate system the susceptibility tensor of
the AF ordered two level system is identical to the same tensor for
the paramagnetic case in the original (x,y,z) hexagonal coordinate
system. It has only nonzero elements for x,y and is given by

\[\tensor{u}'(\omega)=\frac{\alpha^2\la S\ra}{\Delta'^2-\omega^2} 
\left(\matrix{
\Delta'  & i\omega\cr
-i\omega & \Delta'\cr}\right)\]

\begin{equation}
\label{URO}
\tensor{u}^{-1'}(\omega)=\frac{1}{\alpha^2\la S\ra} 
\left(\matrix{
\Delta'  & -i\omega\cr
i\omega & \Delta'\cr}\right)
\end{equation}

This form will be used in the secular equation (\ref{SEC})

\end{document}